\documentclass[12pt]{article}

\pdfoutput=1

\usepackage{graphicx}
\usepackage{epsfig}
\usepackage{amsfonts}
\usepackage{amssymb}
\usepackage{amsmath}
\usepackage{dsfont}
\usepackage{bbm}
\usepackage{cite}
\usepackage{multirow}
\usepackage{colortbl}

\def\VIIA{V_{\textrm{\begin{scriptsize}IIA\end{scriptsize}}}}

\def\Vloc{V_{\textrm{\begin{scriptsize}loc\end{scriptsize}}}}
\def\VNS{V_{\textrm{\begin{scriptsize}NS-NS\end{scriptsize}}}}
\def\VRR{V_{\textrm{\begin{scriptsize}R-R\end{scriptsize}}}}

\def\beq{\begin{equation}}
\def\eeq{\end{equation}}
\def\beqa{\begin{eqnarray}}
\def\eeqa{\end{eqnarray}}

\def\eps{{\epsilon }}

\def\cW{{\mathcal W}} 
\def\cK{{\mathcal K}}

\def\cP{{\mathcal P}}

\def\cZ{{\mathcal Z}}
\def\cz{{\mathcal Z}}
\def\cS{{\mathcal S}}
\def\cT{{\mathcal T}}

\def\fg{{\mathfrak g}}

\def\VIIA{V_{\textrm{\begin{scriptsize}IIA\end{scriptsize}}}}

\def\Vloc{V_{\textrm{\begin{scriptsize}loc\end{scriptsize}}}}

\begin{document}
\pagestyle{plain}

\makeatletter
\@addtoreset{equation}{section}
\makeatother
\renewcommand{\theequation}{\thesection.\arabic{equation}}
\pagestyle{empty}
\rightline{IFT-UAM/CSIC-09-51}
\vspace{5mm}
\begin{center}
\LARGE{\bf
Complete classification of Minkowski vacua in generalised flux models
\\[5mm]}
\large{
Beatriz de Carlos${}^{a}$,  Adolfo Guarino${}^{b}$ and Jes\'us M. Moreno${}^{b}$
\\[3mm]}
\small{
${}^a$
School of Physics and Astronomy, University of Southampton,\\[-0em]
Southampton SO17 1BJ, UK \\[2mm]
${}^b$
Instituto de F\'{\i}sica Te\'orica UAM/CSIC,\\[-0em]
Facultad de Ciencias C-XVI, Universidad Aut\'onoma de Madrid, \\[-0em]
Cantoblanco, 28049 Madrid, Spain\\[2mm]
}
\small{\bf Abstract} \\[3mm]
\end{center}
{\small
We present a complete and systematic analysis of the Minkowski extrema of the ${\cal N}=1$, $D=4$ Supergravity potential  obtained from type II orientifold models that are T-duality invariant, in the presence of generalised fluxes. Based on our previous work on algebras spanned by fluxes, and the so-called  no-go theorems on the existence of Minkowski and/or de Sitter vacua, we perform a partly analytic, partly numerical analysis of the promising cases previously hinted. We find that the models contain Minkowski extrema with one tachyonic direction. Moreover, those models defined by the Supergravity algebra  $\,\mathfrak{so(3,1)^2}\,$ also contain Minkowski/de Sitter minima that are totally stable. All Minkowski solutions, stable or not, interpolate between points in parameter space where one or several of the moduli go to either zero or infinity, the so-called singular points. We finally reinterpret our results in the language of type IIA flux models, in order to show explicitly the contribution of the different sources of potential energy to the extrema found. In particular, the cases of totally stable Minkowski/de Sitter vacua require of the presence of non-geometric fluxes.
}
\vspace{45mm}
\begin{flushleft}
\rule{170mm}{0.5pt}\\
\begin{footnotesize}e-mail: 
b.de-carlos@soton.ac.uk , adolfo.guarino@uam.es , jesus.moreno@uam.es 
\end{footnotesize} \\
\end{flushleft}

\newpage
\setcounter{page}{1}
\pagestyle{plain}
\renewcommand{\thefootnote}{\arabic{footnote}}
\setcounter{footnote}{0}


\tableofcontents

%
%

\section{Objectives, background  and outline}
\label{sec:intro}

The study of moduli stabilisation is a crucial step in establishing a link between low energy phenomenology, which is about to be thoroughly explored at the LHC, and string theory constructions in four dimensions. In particular an understanding of how Supersymmetry (SUSY) breaking happens in this context is mandatory, in order to proceed with this ``top-bottom" approach to linking strings and low energy physics.  

In this paper we continue with the programme already started in ref.~\cite{deCarlos:2009fq}, and we perform a systematic search for moduli vacua within the flux models that looked most promising after scanning them through the so-called no-go theorems~\cite{Hertzberg:2007wc,Silverstein:2007ac,Haque,Caviezel:2008tf,Flauger:2008ad} on the existence of de Sitter vacua. The result is successful and the promising case identified in \cite{deCarlos:2009fq} happens to provide with the already-mentioned de Sitter vacua with all moduli stabilised at reasonable values. With a certain tuning of one of the parameters such vacua can be made Minkowski. The process of searching for these solutions is systematic and could be easily generalised to other models.

In order to explain our results in the most straightforward way in the following sections, let us now recall  the main issues addressed in \cite{deCarlos:2009fq}, and also the main conclusions achieved.

The starting point is the set of  $\,\mathcal{N}=1$, type II  orientifold models that are T-duality invariant and allowed by the symmetries of the  $\,\mathbb{T}^{6}/ (\mathbb{Z}_{2} \times \mathbb{Z}_{2})\,$ isotropic orbifold \cite{STW,Aldazabal:2006up,Font:2008vd,Guarino:2008ik,Aldazabal:2008zza}. Specifically, we concentrate on IIB orientifold models with O3/O7-planes (and, generically, also with D3/D7-branes) in which a background for the non-geometric $\,Q\,$ flux, as well as for the NS-NS $\,\bar{H}_{3}\,$ and R-R $\,\bar{F}_{3}\,$ fluxes, can be consistently switched on. Within these models a classification of all the compatible non-geometric $Q$ flux backgrounds was carried out in ref.~\cite{Font:2008vd}. Subsequently, our previous work \cite{deCarlos:2009fq} extended the results of \cite{Font:2008vd}  to include $\,\bar{H}_{3}\,$ flux, providing a complete classification of the Supergravity algebra $\,\fg\,$, defined by 
\beq
\left[ X^{a},X^{b} \right] = Q_{c}^{ab} \, X^{c}  \hspace{5mm} , \hspace{5mm}
\left[ Z_{a}\,,X^{b} \right] = Q_{a}^{bc} \, Z_{c} \hspace{5mm} , \hspace{5mm}
\left[ Z_{a},Z_{b} \right] = \bar{H}_{abc} \, X^{c}  \ .
\label{IIBalgebra}
\eeq
Here, $\,Z_{a}\,$ and $\,X^{a}$, with $\,a=1,\ldots,6\,$,  are the isometry and gauge generators coming from the reduction of the metric and the $B$-field with fluxes \cite{STW}, respectively. 

Once the allowed fluxes/algebras were defined, we were able to write down a superpotential and, consequently, a scalar potential. We then concentrated on studying the existence of de Sitter (dS) and Minkowski (Mkw) vacua, which are interesting for phenomenology, i.e. that break Supersymmetry. As mentioned above, we made use of some no-go theorems concerning the existence of such vacua,  as well as of the mechanisms proposed to circumvent them \cite{Hertzberg:2007wc,Silverstein:2007ac,Haque,Caviezel:2008tf,Flauger:2008ad}. The subtlety here was that these theorems were mostly proposed in the language of a type IIA generalised flux compactification, including O6-planes and D6-branes. We therefore had to develop a dictionary between the contributions to the scalar potential in the IIA language, in which the no-go theorems were formulated, and the IIB one in which we performed the classification of the Supergravity algebras. By means of this dictionary, we excluded the existence of dS/Mkw vacua in more than {\bf half} of the effective models based on non-semisimple Supergravity algebras. On the other hand, those based on semisimple algebras survive the no-go theorem and stand a chance of having all moduli stabilised.

With the set of effective models that are phenomenologically interesting (aka SUSY breaking ones) narrowed down to a few, we now present a detailed numerical study of potential vacua. In section 2 we define the SUGRA potential and the different models allowed, based on the algebra classification. Section 3 explains the method used to find extrema for this multivariable (6 real fields) potential. The process is analytic in what involves the fields that enter the superpotential linearly (i.e. the dilaton, $\cS$,  and the $\cT$ modulus), whereas it has to be tackled numerically when dealing with the $\cZ$ modulus, which has a cubic dependence. That is done thoroughly in section 4, where we present the main results of the paper, addressing all models one by one. In section 5 we reinterpret our results in the language of type IIA constructions, splitting the potential energy in terms of the different contributions. This is a useful exercise in terms of comparing our results to previous ones in the literature, mainly those addressed to illustrate the no-go theorems. We finally conclude in section 6.

\section{The $\,\mathcal{N}=1\,$ SUGRA models}
\label{sec:SUGRA}

In this section we present a set of  $\,\mathcal{N}=1 \,$ T-duality invariant effective Supergravities. They arise from type IIB generalised flux compactifications on $\,\mathbb{T}^{6}/ (\mathbb{Z}_{2} \times \mathbb{Z}_{2})\,$ isotropic orientifolds with O3/O7-planes \cite{deCarlos:2009fq,STW,Aldazabal:2006up,Font:2008vd,Guarino:2008ik,Aldazabal:2008zza}. The orientifold involution action allows a background for the NS-NS, $\,\bar{H}_{3}\,$, and R-R, $\,\bar{F}_{3}\,$, 3-form fluxes, as well as for the so-called non-geometric, $Q$, tensor flux. These fluxes play a double role at the four dimensional level: on the one hand, they determine the Supergravity algebra $\,\fg\,$ of (\ref{IIBalgebra}), entering it as structure constants and, therefore, being constrained by the Jacobi identities. On the other hand, the fluxes induce a $\,\mathcal{N}=1\,$ superpotential for the moduli fields of the compactification, which can potentially lead to their stabilisation. 

A new approach that combines these two  aspects to explore the phenomenology of generalised flux compactifications was introduced in \cite{Font:2008vd}, and further developed in \cite{deCarlos:2009fq}. It is based on making extensive use of the orbifold symmetries to classify the set of Supergravity algebras embeddable within the fluxes in (\ref{IIBalgebra}), and derive their characteristic flux induced superpotentials. The fluxes entering these superpotentials automatically satisfy all the constraints arising as Jacobi identities of the algebra (\ref{IIBalgebra}). 

The resulting models can be organized according to the $B$-field reduction they are built on. In other words, according to the non-trivial $\,\fg_{gauge}\,$ subalgebra in (\ref{IIBalgebra}) spanned by the gauge $X^{a}$ generators and specified by the non-geometric $\,Q\,$ flux background. The set of such reductions was found to include the semisimple $\,\mathfrak{so(3,1)}\,$ and $\,\mathfrak{so(4)}\,$ algebras, together with the non-semisimple $\,\mathfrak{su(2)} + \mathfrak{u(1)^{3}}\,$, $\,\mathfrak{iso(3)}\,$ and $\,\mathfrak{nil}$, giving rise to five non-equivalent Supergravity models\footnote{The $\,\mathfrak{nil}\,$ algebra was denoted $\,n3.5\,$ in \cite{Grana:2006kf}.}.  They are described by a K\"ahler $\,\cK\,$ potential and a superpotential $\,\cW\,$ of the form
\beq
\begin{array}{ccl}
\mathcal{K} & = &-3 \,\log\left( -i\,(\mathcal{Z}-\bar{\mathcal{Z}})\right)  - 
\,\log\left( -i\,(\mathcal{S}-\bar{\mathcal{S}})\right)  - 3 \,\log\left(- i\,(\mathcal{T}-\bar{\mathcal{T}}) \right) \ ,  \\[2mm]
\mathcal{W}  & = &  |\Gamma|^{3/2} \left[ \, \mathcal{T} \, \cP_3(\cz)  \,  + \, \mathcal{S} \, \cP_2(\cz) \, -\, \xi_{3} \, \tilde{\cP}_{2}(\cZ) \,+\,  \xi_{7} \, \tilde{\cP}_{3}(\cZ) \, \right] \ ,
\label{kwModular}
\end{array}
\eeq
with $\,\cP_{2,3}(\cZ)\,$ being up to cubic polynomials in the $\,\cZ\,$ modulus, and where $\,\tilde{\cP}_{i}(\cZ)\,$ denotes the dual of $\,\cP_{i}(\cZ)\,$ such that $\, \cP_{i} \rightarrow \frac{\tilde{\cP}_{i}}{\cZ^{3}}\,$ when $\,\cZ \rightarrow -\frac{1}{\cZ}$. The set of moduli fields and flux parameters appearing in (\ref{kwModular}) are now introduced following the notation and conventions of \cite{deCarlos:2009fq}.
\\[4mm]
\textbf{Moduli fields} 
\\[1mm]
There is one \textit{R-R shifted} dilaton, $\,\cS\,$, and one \textit{R-R shifted} K\"ahler modulus, $\,\cT\,$, which enter the superpotential linearly. Additionally, there is one \textit{redefined} complex structure modulus, $\,\cZ\,$, which relates to the original complex structure of the compactification via the non-linear action of a $\,\Gamma \in \textrm{GL}(2,\mathbb{R})\,$ matrix. The $\,\cZ\,$ modulus enters the superpotential in eqs~(\ref{kwModular}) through the $\,\cP_{2,3}(\cZ)\,$ and $\,\tilde{\cP}_{2,3}(\cZ)\,$ flux-induced polynomials. Taking the imaginary part of the original complex structure modulus to be positive at any physical vacuum, we will adopt the convention $\,|\Gamma| > 0\,$ without loss of generality. This implies that $\,\textrm{Im}\cZ_{0}  > 0\,$ at any physical vacuum. Moreover, the relation between the moduli values at the vacuum and certain physical quantities, such as the string coupling and the internal volume, imposes that $\,\textrm{Im}\cS_{0} > 0\,$ and $\,\textrm{Im}\cT_{0} > 0 \,$ at the vacuum.
\\[4mm]
\textbf{Couplings} 
\\[1mm]
The expression for $\,\cW\,$ given in eq.~(\ref{kwModular}), contains $\,\cT \cZ^{n}\,$ terms induced by the non-geometric $Q$ tensor flux together with $\,\cS\cZ^{n}\,$ terms induced by the ordinary $\bar{H}_{3}$ flux, up to $\,n=3$. These couplings are totally determined by the form of the $\,\cP_{3}(\cZ)\,$ and $\,\cP_{2}(\cZ)\,$ flux-induced polynomials respectively. 
\begin{table}[h]
\small{
\renewcommand{\arraystretch}{1.15}
\begin{center}
\begin{tabular}{|c|c|c|}
\hline
 $\fg_{gauge}$                             & $\cP_3(\cz)/3$ & $\cP_2(\cz)$ \\
\hline
\hline
 $\mathfrak{so(3,1)}$   & $-\cz^3 - \cz$   &  $\eps_{1} \left( - \cz^{3}  + 3 \, \cz  \right)+ \eps_{2} \left( 1 -  3  \, \cz^{2}   \right)   $   \\
\hline
 $\mathfrak{so(4)}$   & $ \cz^3 - \cz$   &  $\eps_{1} \left( \cz^{3}  + 3 \, \cz  \right)+ \eps_{2} \left( 1 +  3  \, \cz^{2}   \right)   $   \\
\hline
 $\mathfrak{su(2) + u(1)^3}$  & $\cz$ & $ \eps_1 \, \cz^3 + \eps_2 $ \\
\hline
 $\mathfrak{iso(3)}$   & $ -\cz$   &  $    3 \, \eps_{1} \, \cz + \eps_{2}     $   \\
\hline
 $\mathfrak{nil}$   & $  1 $   &  $   \eps_{1}  -  3  \, \eps_{2} \, \cz  $   \\
\hline
\end{tabular}
\end{center}
\caption{The $\cP_3(\cz)$ and $\cP_2(\cz)$ flux-induced polynomials.}
\label{tablecP23}
}
\end{table}

The polynomial $\,\cP_{3}(\cZ)\,$, induced by the Q flux, is totally fixed after the choice of the $B$-field reduction, namely after specifying $\,\fg_{gauge}$. However, the polynomial $\,\cP_{2}(\cZ)\,$, induced by the $\,\bar{H}_{3}\,$ flux, depends on two real parameters $\,(\eps_{1},\eps_{2})\,$ which determine the extension of $\,\fg_{gauge}\,$ to a Supergravity algebra, $\,\fg$. Specific examples of these polynomials are presented in table~\ref{tablecP23}. Finally there are also $\cZ^n$ self-interaction couplings induced by the R-R $\,\bar{F}_{3}\,$ flux. They are specified by two real parameters, $\,(\xi_{3},\xi_{7})\,$, which relate to the number and types of localised sources present, i.e. O3/O7-planes and D3/D7-branes, through the tadpole cancellation conditions \cite{deCarlos:2009fq}.

%
%

\section{Minimisation conditions}

Given the K\"ahler potential and the superpotential of eqs~(\ref{kwModular}), the dynamics of the moduli fields $\Phi\equiv\,\left( \cZ,\cS,\cT\right)\,$ are determined by the standard $\,\mathcal{N}=1\,$ scalar potential
\beq
V = e^\cK \left(  \sum_{\Phi} \cK^{\Phi\bar \Phi} |D_\Phi \cW|^2 - 3|\cW|^2 \right)  \ ,
\label{VModular}
\eeq
where $\,\cK^{\Phi \bar{\Phi}}\,$ is the inverse of the K\"ahler metric $\cK_{\Phi \bar{\Phi}} \equiv \frac{\partial \cK}{\partial \Phi \partial \bar{\Phi}}$ , and $\,D_\Phi \cW = \frac{\partial \cW}{\partial \Phi} + \frac{\partial \cK}{\partial \Phi} \cW\,$ is the K\"ahler derivative. Moduli fields are stabilised at the minimum of the potential energy, taking a vacuum expectation value $\,\Phi_{0}\,$ (VEV) determined by the conditions
\beq
\left. \frac{\partial V}{\partial \Phi} \right|_{\Phi=\Phi_{0}}= 0.
\label{critpoint}
\eeq
From now on, our objective will be to solve the above system  (\ref{critpoint}) of high degree polynomial equations together with the physical requirement of $\,\left. V \right|_{\Phi=\Phi_{0}} \gtrsim 0$, namely, de Sitter (dS), almost Minkowski (Mkw) solutions. Our strategy will consist on finding the exactly Mkw solutions to the minimisation conditions, and then looking for dS extrema continuously connected to them via a deformation of the parameter space.

Since the moduli $\,\mathcal{S}\,$ and $\,\mathcal{T}\,$ enter the superpotential (\ref{kwModular}) linearly, the scalar potential $\,V\,$ computed from (\ref{VModular}) can be written as
\beq
 V = |\Gamma|^{3}  \,  e^{\cK}  \, \Big( \, m_0  + 2 \, m_i \, x_i  +   M_{ij} \, x_i \, x_j \, \Big)  \hspace{5mm}  \mathrm{where} \hspace{5mm}   i , j = 1,..., 4 \ ,
\label{V_quad}
\eeq
and
\beq
x  = \Big( \, \textrm{Re}\cS \,,\,  \textrm{Re}\cT \,,\, \textrm{Im}\cS \,,\, \textrm{Im}\cT  \, \Big) \ .
\label{xvec}
\eeq
Note that, because of the form of the superpotential in eq.~(\ref{kwModular}),  $\,m_0\,$ and $\,m_i\,$ depend on $\,( \, \cZ  \,,\, \epsilon_{1,2}  \,,\,  \xi_{3,7}  \, )\,$,  while the matrix $\,M\,$  does not depend on the R-R flux parameters $\,\xi_{3,7}$. 

The VEVs of the $\,\cS_{0}\,$ and $\,\cT_{0}\,$  moduli that extremise the potential at $ V = 0 $ satisfy 
\beq
\Big( \, \textrm{Re}\cS_0  \,,\, \textrm{Re}\cT_0  \,,\, \textrm{Im}\cS_0  \,,\, \textrm{Im}\cT_0 \, \Big)  = \left. - M^{-1} \, m \right|_{\cZ=\cZ_0} 
\label{min_x} \ ,
\eeq
where  we have assumed a non-degenerate $\,M\,$ matrix. Otherwise there would be flat directions and the stabilisation of $\cS$ and $\cT$ would remain incomplete. It is worth mentioning that, when we plug a particular pair $\left\lbrace \cP_2(\cZ) , \cP_3(\cZ) \right\rbrace$ of polynomials from 
table~\ref{tablecP23},  $M$ becomes box diagonal and splits into  two $2 \times 2$ matrices. In other words, axion and volume moduli do not 
mix\footnote{The  subtle cancellation of the cross terms is a consequence  of the Jacobi identities  of the Supergravity algebra (\ref{IIBalgebra}), in particular  of the $\,\bar{H}_{x[bc}\,Q_{d]}^{ax} = 0\,$ constraints.} in the quadratic polynomial of eq.~(\ref{V_quad}).

Using eq.~(\ref{min_x})  the $V=0 $ condition reads
\beq
m_0  -   M_{ij}^{-1}    m_i  m_j  = 0  \ ,
\label{min_0}
\eeq
and provides us with the first constraint between the  $\cz$ modulus and the $\,\epsilon_{1,2}\,$ and $\,\xi_{3,7}\,$ parameters at the Mkw vacua.  
The function appearing in eq.~(\ref{min_0}),
\beq
\mathbb{V}(\cZ) \equiv m_0  -   M_{ij}^{-1}    m_i  m_j  \ ,
\label{v_eff}
\eeq
plays an important role in the calculation. The  equations  derived from $\,\partial_{\textrm{Re}\cZ }V  = \partial_{\textrm{Im}\cZ} V = 0\,$ are just
\beq
\partial_{\textrm{Re}\cZ} \,  \mathbb{V}  =  0 \hspace{10mm} \textrm{ and } \hspace{10mm}  \partial_{\textrm{Im}\cZ } \mathbb{V}  = 0 \ ,
\label{min_12}
\eeq
where again we have used $V=0$ and  eq.~(\ref{min_x}). The reduced potential, $\,\mathbb{V}(\cZ)\,$, captures the Mkw extrema of  $V$ and some of their stability properties. In particular, tachyonic Mkw extrema in $\,\mathbb{V}(\cZ)\,$ have their origin in tachyonic Mkw extrema of the full potential $V$.

\subsection{An example: the $\,\mathfrak{nil}\,$ based models}
\label{sec:Nil}

We now clarify the previous procedure by explaining the $\,\mathfrak{nil}\,$ case in detail. This algebra is defined by the superpotential \cite{deCarlos:2009fq}
\beq
\begin{array}{ccl}
\mathcal{W}  & = &  |\Gamma|^{3/2} \left[ 3 \, \mathcal{T} + \mathcal{S}  \left( \eps_{1}  -  3  \, \eps_{2} \, \cz  \right) - \xi_{3} \left( \eps_{1} \, \cZ^3 + 3  \, \eps_{2} \, \cz^2 \right) + 3 \,\xi_{7} \, \cZ^3 \right] \ ,
\end{array}
\label{WNil}
\eeq
and the K\"ahler potential in (\ref{kwModular}). 

The function $m_0$, derived from this superpotential, is given by~\footnote{To make the expressions lighter we replace $\,(\textrm{Im}\Phi)^{q}$ with $\,\textrm{Im}\Phi^{q}\,$, and similarly for any powers of $\textrm{Re}\Phi$.}
\beqa
m_{0} &=& 4 \, |\cZ|^{2} 
\left[   
\Big( \left( \eps_{1}\, |\cZ|^{2} + 3 \, \eps_{2} \, \textrm{Re}\cZ \right) \, \xi_{3} - 3 \, |\cZ|^{2} \, \xi_{7} \Big)^2 
+ 3 \, \epsilon_{2}^{2} \, \xi_{3}^{2}  \, \textrm{Im}\cZ^{2}   
\right] \ ,
\label{m0Nil}
\eeqa
whereas the functions $m_i$ are
\beq
\begin{array}{lcl}
\\[-4mm]
m_{1} & = &   4 \,  \textrm{Re}\cZ  
\left[  \textrm{Re}\cZ \,  \Big( 3 \, \eps_{2} \, \textrm{Re}\cZ  -\eps_{1} \Big) 
\Big( \textrm{Re}\cZ \, ( \eps_{1} \, \xi_3 - 3 \, \xi_{7}  ) +  3 \, \eps_{2} \, \xi_3  \Big) \, + \right. \\
      &   & \phantom{4  \,  \textrm{Re}\cZ \;} \left.
3 \, \eps_2 \, \textrm{Im}\cZ^2 \, \Big( \textrm{Re}\cZ \, ( \eps_{1} \, \xi_3 - 3 \, \xi_{7}  ) +  2 \, \eps_{2} \, \xi_3   \Big)  \right]  \ , \\[2mm]

m_{2} & = & -12 \, \textrm{Re}\cZ^{2}  \left[ \, \textrm{Re}\cZ \, ( \eps_{1} \, \xi_3 - 3 \, \xi_{7}  ) +  3 \, \eps_{2} \, \xi_3 \, \right]  \ ,  \\[2mm]

m_{3} &= &-4 \, \textrm{Im}\cZ^{3} \left[ \, \eps_{1} \, \left(  \eps_{1} \, \xi_{3} - 3 \, \xi_{7} \right)  + 3 \, \eps_{2}^2  \, \xi_{3} \, \right]       \ ,  \\[2mm]

m_{4} & =&  -12 \, \textrm{Im}\cZ^{3}  \left(  \eps_{1} \, \xi_{3} - 3 \, \xi_{7} \right)  \ . \\[2mm]
\end{array}
\label{mNil}
\eeq
As mentioned above, the $4 \times 4$ symmetric matrix $\,M\,$ splits into two $2 \times 2$ matrices, the first one acting on the  axions $\, \textrm{Re}\cS\,$ and $\,\textrm{Re}\cT \,$ with 
\beq
\begin{array}{lcl}
M_{11} & = &  4  \, \left( 3 \, \eps_{2} \, \textrm{Re}\cZ  - \eps_{1} \right) ^2 +  12 \, \eps_{2}^2  \, \textrm{Im}\cZ^2   \ , \\
M_{22} & = & 36 \ ,\\
M_{12} & = & -12 \,  \left(  3 \, \eps_{2} \, \textrm{Re}\cZ  -\eps_{1} \right)  \ , 
\end{array}
\label{M_axi_Nil}
\eeq
and the second one on the volumes  $\,\textrm{Im}\cS \,$ and $\,\textrm{Im}\cT\,$ with
\beq
\begin{array}{lcl} 
M_{33} & = &   4  \, \left( 3 \, \eps_{2} \, \textrm{Re}\cZ  -\eps_{1} \right) ^2 +  12 \, \eps_{2}^2  \, \textrm{Im}\cZ^2     \ , \\
M_{44} & = & 12 \  ,\\
M_{34} & = & 0  \ . 
\end{array}
\label{M_vol_Nil}
\eeq
The absence of flat directions implies $\,\eps_{2} \neq 0\,$. Otherwise, only the linear combination $\,3\, \cT + \eps_{1}\, \cS\,$ enters the superpotential (\ref{WNil}), and the axionic part of its orthogonal combination cannot be fixed.

At this stage, we do not  know yet if there will be full, stable Mkw minima. If any, the axions of $\cS$ and $\cT$ will be  fixed at the values
\beq
\begin{array}{rcl}
\epsilon_2 \, \textrm{Re}\cS_{0} &=& - \textrm{Re}\cZ_{0}^2 \left( \epsilon_1 \, \xi_3 - 3 \, \xi_7 \right) - 2 \, \textrm{Re}\cZ_{0} \, \epsilon _2 \, \xi_3 \ , \\[2mm]
3 \, \textrm{Re}\cT_{0} &=& - \textrm{Re}\cS_{0} \, \epsilon_1 + 3 \, \textrm{Re}\cS_{0} \, \textrm{Re}\cZ_{0} \, \epsilon_2 + \textrm{Re}\cZ_{0}^3 \left( \epsilon_1 \, \xi_3 - 3 \, \xi_7  \right) + 3 \, \textrm{Re}\cZ_{0}^2 \, \epsilon_2 \, \xi_3 \ ,
\end{array}
\eeq
while their volume partners will be given by
\beq
\textrm{Im}\cS_{0} =  \left. - \frac{ m_3 }{ M_{33}   }   \right|_{\cZ=\cZ_{0}} 
\hspace{5mm} , \hspace{5mm}
\textrm{Im}\cT_{0} = \left. - \frac{  m_4}{ M_{44}   }  \right|_{\cZ=\cZ_{0}} \ .
\label{ST_example}
\eeq

Finally we analyse the $\cZ$ modulus stabilisation at Mkw vacua, described by the reduced potential $\,\mathbb{V}(\cZ)\,$ in eq.~(\ref{v_eff}). The physical Mkw extrema conditions require both
\beq
\left\lbrace  \,\, \mathbb{V}  \hspace{2mm} , \hspace{2mm}    \partial_{\textrm{Re}\cZ} \, \mathbb{V}  \hspace{2mm} , \hspace{2mm}    \partial_{\textrm{Im}\cZ} \, \mathbb{V}  \,\, \right\rbrace_{\cZ=\cZ_{0}}  =  0 
\label{idealI}
\eeq
and
\beq
\left\lbrace  \,\, \det{M}   \hspace{2mm} , \hspace{2mm}   M_{44} \,\, m_3 \hspace{2mm} , \hspace{2mm}  M_{33} \,\,  m_4  \,\, \right\rbrace_{\cZ=\cZ_{0}}  \neq  0 \ .
\label{idealJ}
\eeq
The last three conditions ensure a complete stabilisation of $\cS$ and  $\cT$ at non-vanishing  $ \textrm{Im}\cS_{0}$ and $\textrm{Im}\cT_{0}$ values.
Plugging the above expressions for $\,\left( m_0 \,, m_i \, , M \right)$, it can be shown that these two condition sets are incompatible.  Hence we can conclude that there are no Mkw extrema in the Supergravity models based on the $\,\mathfrak{nil}\,$ $B$-field reduction.

%
%

\section{Numerical analysis}
\label{sec:Numerical}

In this section we perform a detailed search of Minkowski extrema for the set of Supergravity models based on the non-semisimple $\,\mathfrak{iso(3)}\,$ and $\,\mathfrak{su(2)} + \mathfrak{u(1)}^{3}\,$, as well as the semisimple $\,\mathfrak{so(4)}\,$ and $\,\mathfrak{so(3,1)}\,$ $B$-field reductions introduced in section~\ref{sec:SUGRA}. The task will be that of solving the set (\ref{idealI}) of polynomial equations 
\beq
\left. \mathbb{V} \right|_{\cZ=\cZ_{0}} = 0 \hspace{10mm}  , \hspace{10mm}   \left.  \frac{\partial \, \mathbb{V}}{\textrm{Re}\cZ}\right|_{\cZ=\cZ_{0}} = 0 \hspace{10mm}  ,  \hspace{10mm}  \left.\frac{\partial \, \mathbb{V}}{\textrm{Im}\cZ}\right|_{\cZ=\cZ_{0}} = 0  \ .
\label{system}
\eeq
The method we will use to find the solutions of (\ref{system}) makes use of the symmetries and the scaling properties of the Supergravity models which are now introduced.

\subsection{Parameter space, discrete symmetries and strategy}

It is worth noticing that the form of the superpotential in eqs~(\ref{kwModular}), in particular that of the polynomials  $\,\cP_{2}(\cZ)\,$ in table~\ref{tablecP23}, allows us to remove the factor $\,|\eps|\equiv \sqrt{\eps_{1}^2 + \eps_{2}^2}\,$ from it (provided it is non-zero) by a rescaling of the $\,\cS\,$ modulus and a redefinition of the $\,\xi_{3}\,$ parameter. Therefore, the angle $\,\tan \theta_{\eps} \equiv   \displaystyle \frac{\eps_{2}}{\eps_{1}} \, $ is the only free parameter coming from the NS-NS flux. Analogously, the (non-vanishing) combination $\,|\xi|\equiv \sqrt{ |\eps|^{2} \,\xi_{3}^{2} + \xi_{7}^{2}}\,$ can be globally factorised in the superpotential by rescaling both the $\,\cS\,$ and  $\,\cT\,$ moduli. This leaves the angle given by $\, \tan \theta_{\xi} \equiv  \displaystyle  \frac{\xi_{7}}{|\eps| \, \xi_{3}} \,$ as the free parameter coming from the R-R flux. These parameter redefinitions and moduli rescalings are given by
\beq
\epsilon_{1} \rightarrow |\epsilon| \cos\theta_{\epsilon} \hspace{4mm},\hspace{4mm}
\epsilon_{2} \rightarrow |\epsilon| \sin\theta_{\epsilon} \hspace{4mm},\hspace{4mm}
\xi_{3} \rightarrow \frac{|\xi|}{|\epsilon|} \cos\theta_{\xi} \hspace{4mm},\hspace{4mm}
\xi_{7} \rightarrow |\xi| \sin\theta_{\xi} \ ,
\label{red_param}
\eeq
together with
\beq
\cS \rightarrow \frac{\cS \, |\xi|}{|\eps|} \hspace{10mm} \textrm{and} \hspace{10mm}  
\cT \rightarrow \cT \, |\xi| \ ,
\label{red_moduli}
\eeq
generating a global factor in the superpotential and, therefore, also in the scalar potential,
\beq
\cW \rightarrow |\Gamma|^{\frac{3}{2}} \, |\xi| \,\,\, \cW\left(\Phi \,;\, \theta_{\eps},\theta_{\xi}\right) \hspace{5mm} \textrm{and} \hspace{5mm} V \rightarrow \frac{|\Gamma|^3 \, |\eps|}{|\xi|^{2}} \,\,\, V\left(\Phi \,;\, \theta_{\eps},\theta_{\xi}\right) \nonumber \ .
\label{global_rescaling}
\eeq
The moduli rescaling in (\ref{red_moduli}) also implies a rescaling of the F-term for all the moduli fields
\beq
F_{\cZ} \rightarrow |\Gamma|^{\frac{3}{2}}\,|\xi|\,F_{\cZ}\left(\Phi \,;\, \theta_{\eps},\theta_{\xi}\right)	  \hspace{2mm},\hspace{2mm}  F_{\cS} \rightarrow |\Gamma|^{\frac{3}{2}}\,|\eps|\,F_{\cS}\left(\Phi \,;\, \theta_{\eps},\theta_{\xi}\right) \hspace{2mm} , \hspace{2mm} F_{\cT} \rightarrow |\Gamma|^{\frac{3}{2}}\,F_{\cT}\left(\Phi \,;\, \theta_{\eps},\theta_{\xi}\right) \ ,
\label{F-term_rescaling}
\eeq
where $\,F_{\Phi}\equiv D_{\Phi} \cW\,$ in eq.~(\ref{VModular}). Then, at any non-supersymmetric extremum with $F_{\Phi=\cZ,\cS,\cT} \neq 0$, Supersymmetry will be mostly broken by $F_\cS$ ($F_\cZ$) when the $|\eps|$ ($|\xi|$) parameter is large, and also by $F_\cT$ when both $|\eps|$ and $|\xi|$ are small. Furthermore, the normalised moduli masses are also sensitive to these rescalings. Specifically, by varying  $\,|\epsilon|\,$, the eigenvectors of the mass matrix are modified. From now on, we will always take $\,|\epsilon|\,$ and $\,|\xi|\,$ to be $\,+1\,$ when presenting numerical examples of moduli masses at an extremum of the potential.

After applying (\ref{red_param}) and (\ref{red_moduli}), the parameter space of the Supergravity models can be understood as a 2-torus with coordinates $\,(\theta_{\eps} , \theta_{\xi})$. These effective models come up with a set of discrete symmetries which allows us to map non-physical solutions into physical ones and \textit{vice versa}. The set of such symmetries act on the moduli fields and the parameter space as follows:
\begin{enumerate}
\item[$i)$]  $\,\cW\,$ is invariant under
\beq
\begin{array}{rclcrcl}
\cS         &  \rightarrow  & - \cS  & \hspace{5mm},& \hspace{5mm} \left( \, \theta_{\epsilon} \,,\, \theta_{\xi} \, \right)      & \rightarrow  & \left( \, \theta_{\epsilon}+\pi \,,\, \pi - \theta_{\xi} \, \right)    \ .
\label{transWS}
\end{array}
\eeq

\item[$ii)$]  $\,\cW\,$ goes to  $-\cW\,$ under these two transformations:
\beq
\begin{array}{rclcrcl}
 \cT        &  \rightarrow  &   -\cT &\hspace{5mm},&\hspace{5mm} \left( \, \theta_{\epsilon} \,,\, \theta_{\xi} \, \right)       & \rightarrow  & \left(\, \theta_{\epsilon}+\pi  \,,\, 2 \, \pi - \theta_{\xi}  \, \right) \ .   
\label{transWT}
\end{array}
\eeq
\beq
\begin{array}{rclcrcl}
\cZ & \rightarrow &  -  \cZ &\hspace{5mm},&\hspace{5mm}  \left( \, \theta_{\epsilon} \,,\, \theta_{\xi} \, \right)    & \rightarrow &    \left(\, 2 \, \pi - \theta_{\epsilon}  \,,\, \theta_{\xi}+\pi \,\right) \ . 
\label{transWZ}
\end{array}
\eeq

\item[$iii)$] Finally, since the parameters entering the superpotential are real, we can combine field conjugation with the above transformations to obtain an additional symmetry
\beq
\begin{array}{rclcrcl}
\left( \, {\cZ \,,\, \cS \,,\, \cT} \, \right) & \rightarrow &  -  \left( \, {\cZ \,,\, \cS \,,\, \cT} \, \right)^{*} &\hspace{5mm},&\hspace{5mm} \left( \, \theta_{\epsilon} \,,\, \theta_{\xi} \, \right)    & \rightarrow &    \left( \, 2\, \pi - \theta_{\epsilon} \,,\, \theta_{\xi} \, \right) \ ,
\label{transWAll}
\end{array}
\eeq
which relates physical extrema at $\,\pm  \theta_{\epsilon}$.
\end{enumerate}
These symmetries of the Supergravity models will be extensively used when scanning the parameter space looking for the physical solutions ($\textrm{Im}\Phi_{0} > 0$) to the system (\ref{critpoint}). 
\\[-5mm]

The strategy to perform such a search will be the following: our scanning parameter is the angle $\,\theta_{\eps}$, which needs to be evaluated only in the interval $\,\theta_{\eps} \in [0, \pi]\,$ because of the symmetry (\ref{transWAll}). The value of $\,\theta_{\xi}\,$ can be obtained from the first equation in (\ref{system}) since $\,\tan \theta_{\xi}\,$ enters it quadratically. Substituting $\,\theta_{\xi}(\theta_{\eps},\cZ_{0})\,$ into the original system (\ref{system}), it reduces to
\beq
\left.\frac{\partial \, \mathbb{V}}{\textrm{Re}\cZ}\right|_{\cZ=\cZ_{0}} = h_{1}(\theta_{\eps},\cZ_{0})=0 \hspace{10mm}  \textrm{and}  \hspace{10mm}  \left.\frac{\partial \, \mathbb{V}}{\textrm{Im}\cZ}\right|_{\cZ=\cZ_{0}} = h_{2}(\theta_{\eps},\cZ_{0})=0 \ ,
\label{system_red}
\eeq
where $\,h_{1}\,$ and $\,h_{2}\,$ are complicated functions depending on the Supergravity model under consideration. Provided a value for the angle $\,\theta_{\eps}$, the VEV of $\,\cZ_{0}\,$ can be numerically computed from (\ref{system_red}). After that, and using the value obtained for $\,\theta_{\xi}(\theta_{\eps},\cZ_{0})\,$, the VEVs for the moduli fields $\,\cS\,$ and $\,\cT\,$ can be obtained from (\ref{min_x}). 

In this sense, the modulus $\,\cZ\,$ is the key field in the stabilisation process, whereas  $\,\cS\,$ and $\,\cT\,$ simply get adjusted to generate the extremum of the potential. However, there are singular points given by $\,\textrm{Im}\cZ_{0} = 0$. We find that the value of the $\,\theta_{\eps}\,$ parameter and the VEV of the $\,\textrm{Re}\cZ\,$ modulus at such points can be obtained\footnote{Notice that these conditions correspond to the stabilisation of the $\cS$ and $\cT$ moduli at a globally supersymmetric extremum, namely $\,\partial_{\cS}\cW = \partial_{\cT}\cW = 0$.}  from $\,\cP_{2}(\cZ_{0})=\cP_{3}(\cZ_{0})=0$.

\subsection{Models based on non-semisimple $B$-field reductions}
\label{sec:non-semisimple_models}

The first Supergravity models we will deal with are those based on non-semisimple $B$-field reductions, namely, the $\,\mathfrak{iso(3)}\,$ and the $\,\mathfrak{su(2)} + \mathfrak{u(1)^{3}}\,$ reductions. These models exhibit a special feature: the functions $\,h_{1}\,$ and $\,h_{2}\,$ in (\ref{system_red}) become homogeneous functions, so the set of Mkw extrema for these models has a scaling nature,
\beq
\cZ_{0}(\theta_{\eps}) \propto  | \tan\theta_{\eps}|^{n} \ .
\label{Z_scaling}
\eeq

\subsubsection*{The $\,\mathfrak{iso(3)}\,$ models}
\label{sec:iso3}

\begin{figure}[h!]
\centering
\includegraphics[width=8.5cm, height=6cm]{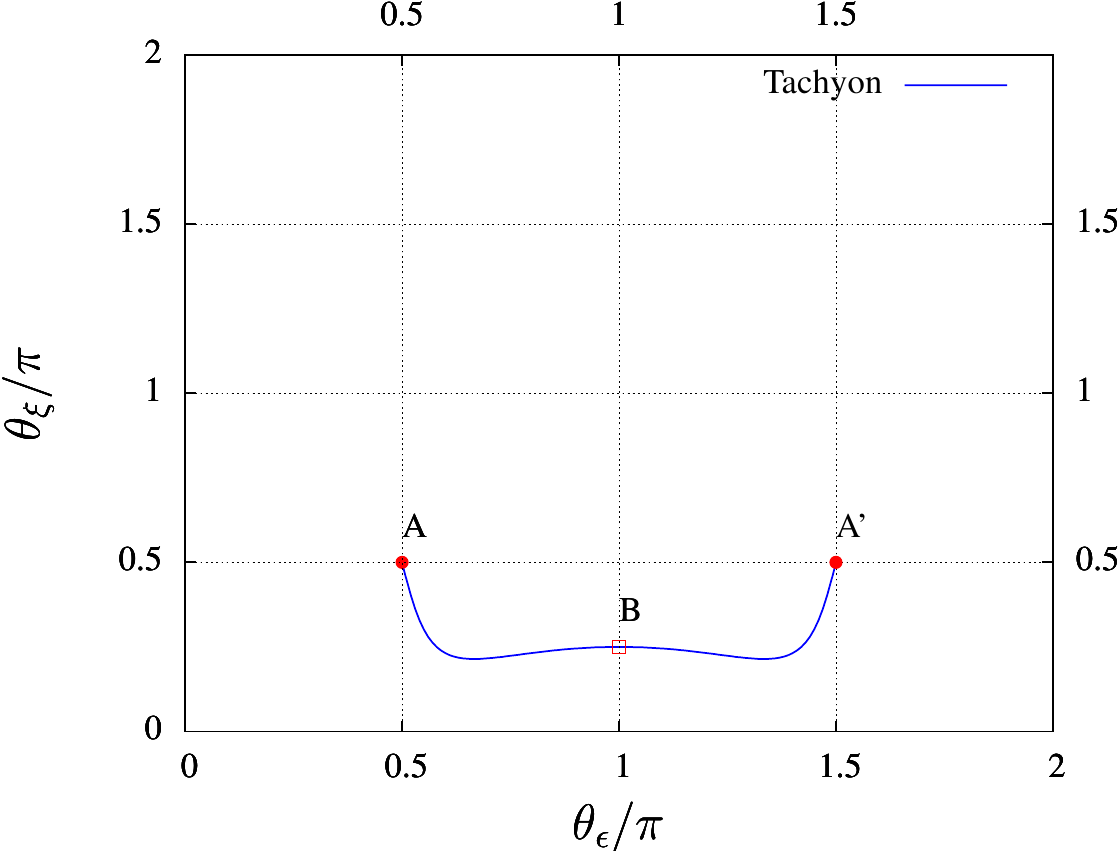}
\includegraphics[width=8.1cm, height=5.7cm]{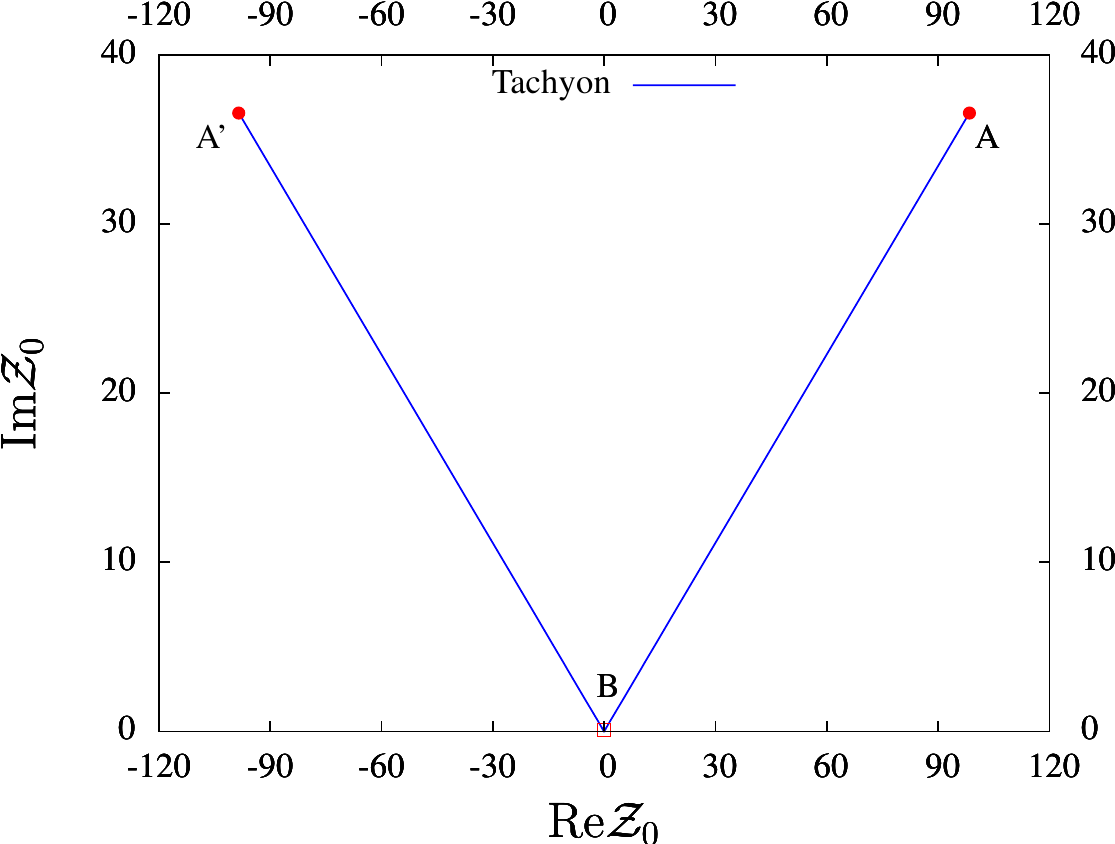}
\caption{Left: location of the Mkw solutions within the parameter space for the Supergravity models based on the $\,\mathfrak{iso(3)}\,$ $B$-field reduction, highlighting the singular points. Right: the set of VEVs of the modulus $\,\cZ\,$, reflecting its scaling nature. The points A and A' correspond to a singular limit $\,|\cZ_{0}| \rightarrow \infty$.}
\label{fig:iso}
\end{figure}

Let us start by exploring  Minkowski solutions for the Supergravity model based on the $\,\mathfrak{iso(3)}\,$ non-semisimple $B$-field reduction. This model is specified by the K\"ahler potential in eqs~(\ref{kwModular}) and the superpotential
\beq
\begin{array}{ccl}
\mathcal{W}  & = &  |\Gamma|^{3/2} \left[-3 \, \mathcal{T} \, \cz \,+ \, \mathcal{S} \, \left( 3 \, \eps_{1} \, \cz + \eps_{2} \right) - \xi_{3} \, \left( \eps_{2} \, \cZ^3 - 3 \, \eps_{1} \, \cz^2 \right)  + 3 \,  \xi_{7} \, \cZ^2 \right] \ .
\end{array}
\label{WIso}
\eeq
Using the procedure introduced in the previous section, we find Mkw extrema in the $\, \epsilon_1 < 0 \,$ range, as shown in figure~\ref{fig:iso}. They are all rescaled solutions of the form 
\beq
\cZ_{0}(\theta_{\eps})=    | \tan\theta_{\eps}| \, (\pm 0.30920 + 0.11495  \, i) \ , 
\eeq
and have a tachyonic direction, hence being unstable. 

The set of singular points in the figure, as well as the Supergravity algebras underlying the different regions in the plots, are summarized as follows:  
\begin{itemize}
\item[$i)$] Points A and A' have an underlying $\,\fg=\mathfrak{iso(3)} \,\oplus_{{\mathbb{Z}_{3}}}\, \mathfrak{u(1)^{6}}\,$ and are conjugate points with respect to the transformations (\ref{transWS}) and (\ref{transWAll}). As we flow towards them, the tachyon aligns with the Im$\cS$ modulus direction and $\,|\Phi_{0}|\rightarrow \infty\,$ for all the moduli fields. Due to their underlying Supergravity algebra, these points were excluded to have dS/Mkw extrema in ref.~\cite{deCarlos:2009fq}.
\\[2mm]
In the following, we will generically refer to such  points as points of excluded Supergravity algebras. They will show up as singularities in the moduli VEVs.

\item[$ii)$] All along the $\,\overline{\textrm{AA'}}\,$ line, including point B located at $\,(\theta_{\eps},\theta_{\xi})=(\pi,\frac{\pi}{4})\,$, there is a unique underlying Supergravity algebra $\,\fg=\mathfrak{so(4)} \,\oplus_{{\mathbb{Z}_{3}}}\, \mathfrak{u(1)^{6}}\,$. As we flow towards this point B, the tachyon aligns with the Im$\cZ$ modulus direction, and $\,|\Phi_{0}|\rightarrow 0\,$ for all the moduli fields, becoming again a singularity in the moduli VEVs.
\\[2mm]
Unlike the previous A and A' points, the Supergravity algebra underlying the point B is not excluded to have dS/Mkw extrema \cite{deCarlos:2009fq}. Therefore, with some abuse of the language, we will refer to these points as dynamical singularities in the moduli VEVs. Observe that the $\,\overline{\textrm{AA'}}\,$ line in the left plot of figure~\ref{fig:iso} is smooth at the singular point B.
\end{itemize}

\subsubsection*{The $\,\mathfrak{su(2) + u(1)^{3}}\,$ models}
\label{sec:su2xu1}

Let us continue with the second set of Supergravity models based on a non-semisimple $B$-field reduction. Those models are based on the $\,\mathfrak{su(2) + u(1)^{3}}\,$ reduction. They are defined by (\ref{kwModular}) with the superpotential 
\beq
\begin{array}{ccl}
\mathcal{W}  & = &  |\Gamma|^{3/2} \left[ 3 \, \mathcal{T} \, \cz \,+ \, \mathcal{S} \, \left(\eps_1 \, \cz^3 + \eps_2 \right) + \xi_{3} \left(\eps_1 - \eps_2 \, \cZ^3 \right) -3 \,  \xi_{7} \, \cZ^2 \right] \ .
\end{array}
\label{WDirect}
\eeq
The set of Minkowski solutions for this model is very similar to that previously analysed. This time, they correspond to solutions of the form
\beq
\cZ_{0}(\theta_{\eps})=    | \tan\theta_{\eps}|^{\frac{1}{3}} \, (\pm 0.99368 + 0.55061  \, i) \ , 
\eeq
and also have a tachyonic direction, being unstable. 
\begin{figure}[h!]
\centering
\includegraphics[width=8.5cm, height=6cm]{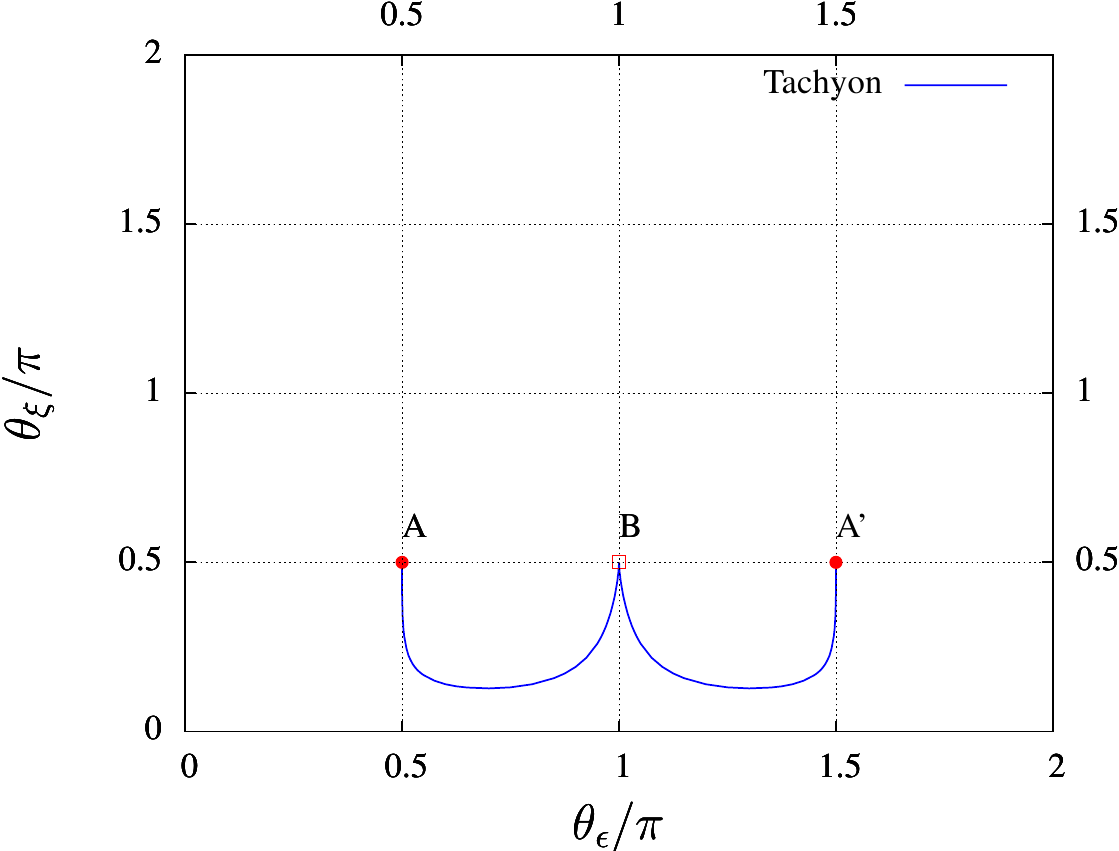}
\includegraphics[width=8.1cm, height=5.7cm]{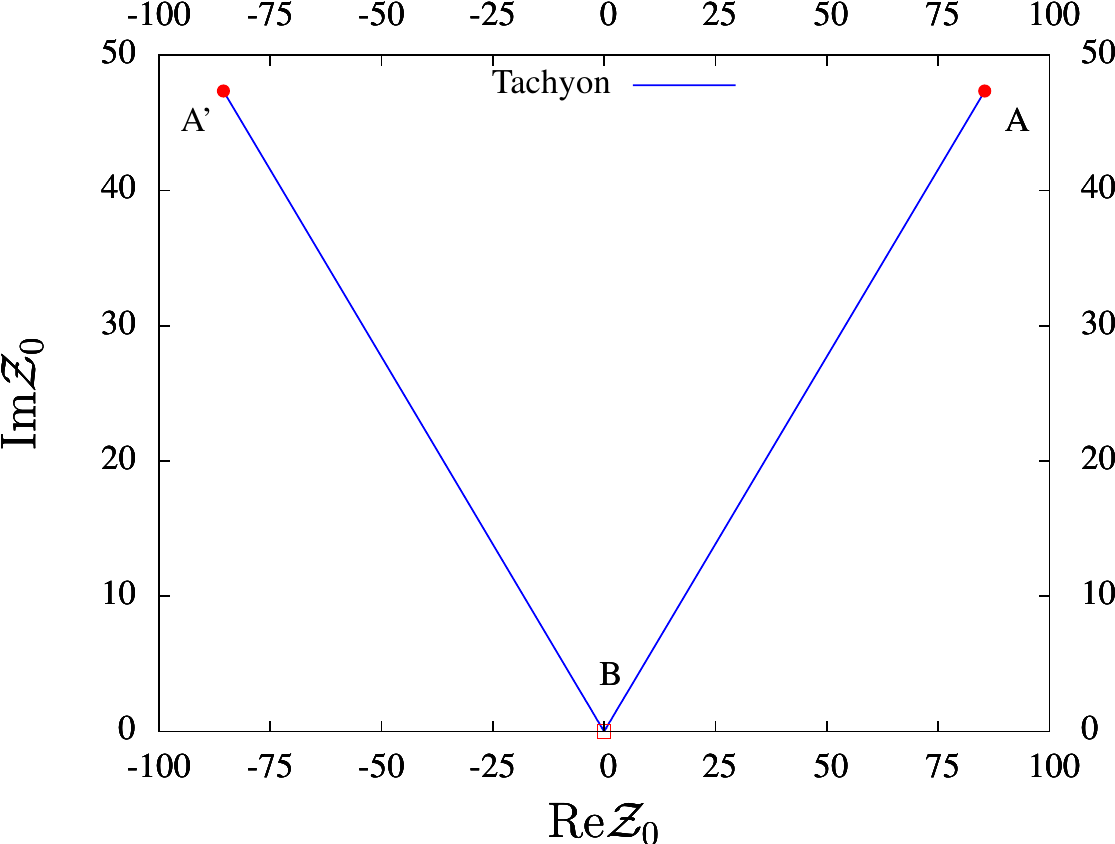}
\caption{Left: location of the Mkw solutions within the parameter space for the Supergravity models based on the $\,\mathfrak{su(2) + u(1)^{3}}\,$ $B$-field reduction, highlighting the singular points. Right: set of VEVs of the modulus $\,\cZ\,$, reflecting its scaling nature. Again, the points A and A' are singular since $\,|\cZ_{0}|\rightarrow \infty$.}
\label{fig:Direct}
\end{figure}

The results are shown in the two plots of figure~\ref{fig:Direct}, where the singular points can be described as follows: 
\begin{itemize}
\item[$i)$] Points A and A' have an underlying $\,\fg=\mathfrak{iso(3)} \,+\, \mathfrak{nil}\,$ and, as in the previous case, are conjugate points with respect to the transformations (\ref{transWS}) and (\ref{transWAll}). As we flow towards these points, the tachyon aligns with the Im$\cS$ modulus direction and also $\,|\Phi_{0}|\rightarrow \infty\,$ for all the moduli fields. Therefore they are again singularities associated to points of excluded Supergravity algebras \cite{deCarlos:2009fq}. 

\item[$ii)$] Along the $\,\overline{\textrm{AB}}\,$ and $\,\overline{\textrm{BA'}}\,$ lines, the Supergravity algebra is $\,\fg=\mathfrak{so(4)} \,+\, \mathfrak{nil}$. However, this time point B corresponds to a different algebra, $\,\fg=\mathfrak{so(4)} \,+\, \mathfrak{u(1)^{6}}\,$, which cannot have Minkowski extrema \cite{deCarlos:2009fq}. As we flow towards this point B, $\,|\Phi_{0}|\rightarrow 0\,$ for all the moduli fields, resulting in a singularity in the moduli VEVs. Observe that the line of Mkw extrema is no longer smooth at this point, around which the tachyon aligns itself along the Im$\cS$ modulus direction.
\end{itemize}

\subsection{Models based on semisimple $B$-field reductions}
\label{sec:semisimple_models}

In the final part of this section we concentrate on the Supergravity models based on the semisimple $B$-field reductions of $\,\mathfrak{so(4)}\,$ and $\,\mathfrak{so(3,1)}$. Their distribution of Minkowski extrema is more involved than that of the previous models based on non-semisimple reductions. This is mainly because the scaling property (\ref{Z_scaling}) no longer takes place. 

As we will see, the distribution of Minkowski extrema draws closed curves in both the parameter space and the $\,\cZ_{0}\,$ complex plane. Although the former has to be understood as a closed curve up to some of the  discrete transformation in (\ref{transWS}) and (\ref{transWT}), the latter is a truly closed curve in the $\,\cZ_{0}\,$ complex plane.

\subsubsection*{The $\,\mathfrak{so(4)}\,$ models}
\label{sec:so4}

The first Supergravity model based on a semisimple $B$-field reduction we are going to describe is that of the $\,\mathfrak{so(4)}\,$ reduction. This model is defined in eqs~(\ref{kwModular}) with the superpotential given by
\beq
\begin{array}{ccl}
\mathcal{W}  & = &  |\Gamma|^{3/2} \left[3\, \mathcal{T} \, \left( \cz^3 - \cz \right) \,+ \, \mathcal{S} \, \left(    \eps_{1} \, \cz^{3} +  3  \, \eps_{2} \, \cz^{2} + 3 \, \eps_{1} \, \cz  + \eps_{2} \right) + \right. \\
&+&  \left. \xi_{3} \left( \eps_{1} -  3  \, \eps_{2} \, \cz  + 3 \, \eps_{1} \, \cz^2 - \eps_{2} \, \cz^3  \right)  - 3 \,  \xi_{7} \, \left( 1 - \cz^2 \right) \right] \ .
\end{array}
\label{WSO4}
\eeq
As it happens for the Supergravity models studied so far, there are only Minkowski solutions with a tachyonic direction. These unstable Mkw solutions are shown in figure~\ref{fig:SO4}, where the singular points highlighted in the plots are now explained:  

\begin{figure}[h!]
\centering
\includegraphics[width=8.5cm, height=6cm]{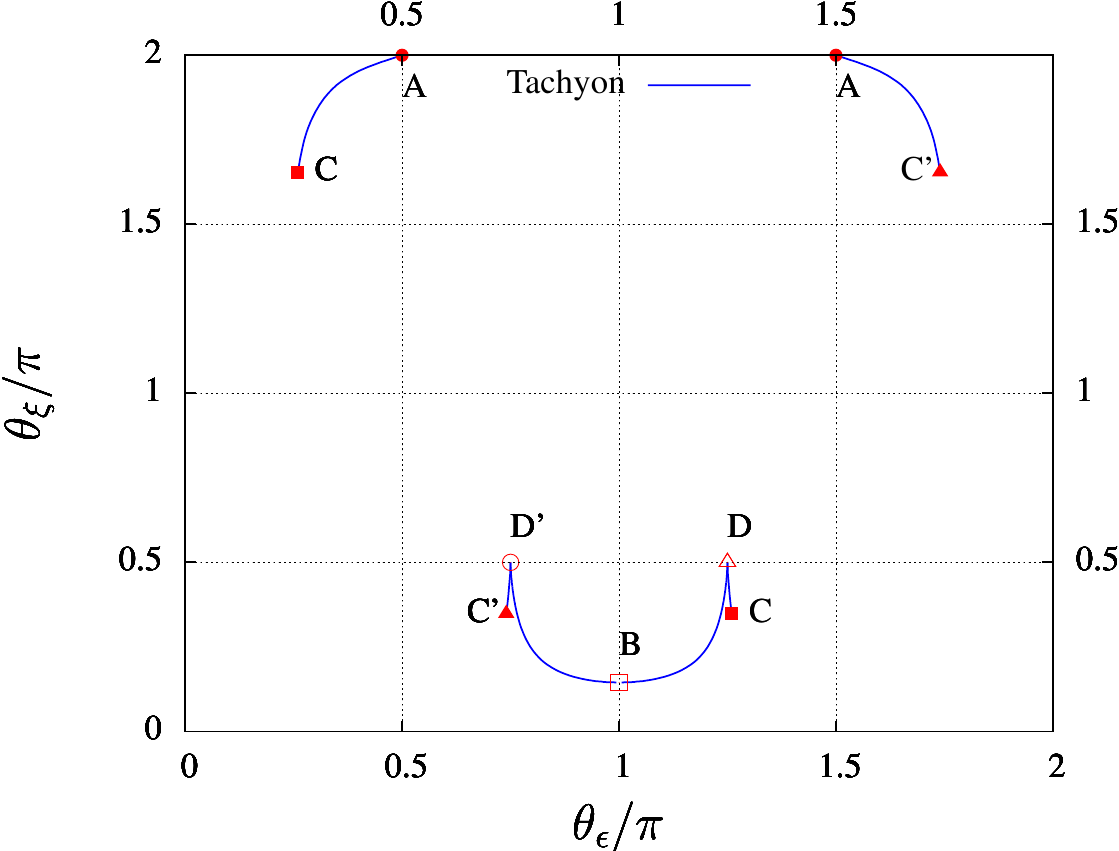}
\includegraphics[width=8.1cm, height=5.7cm]{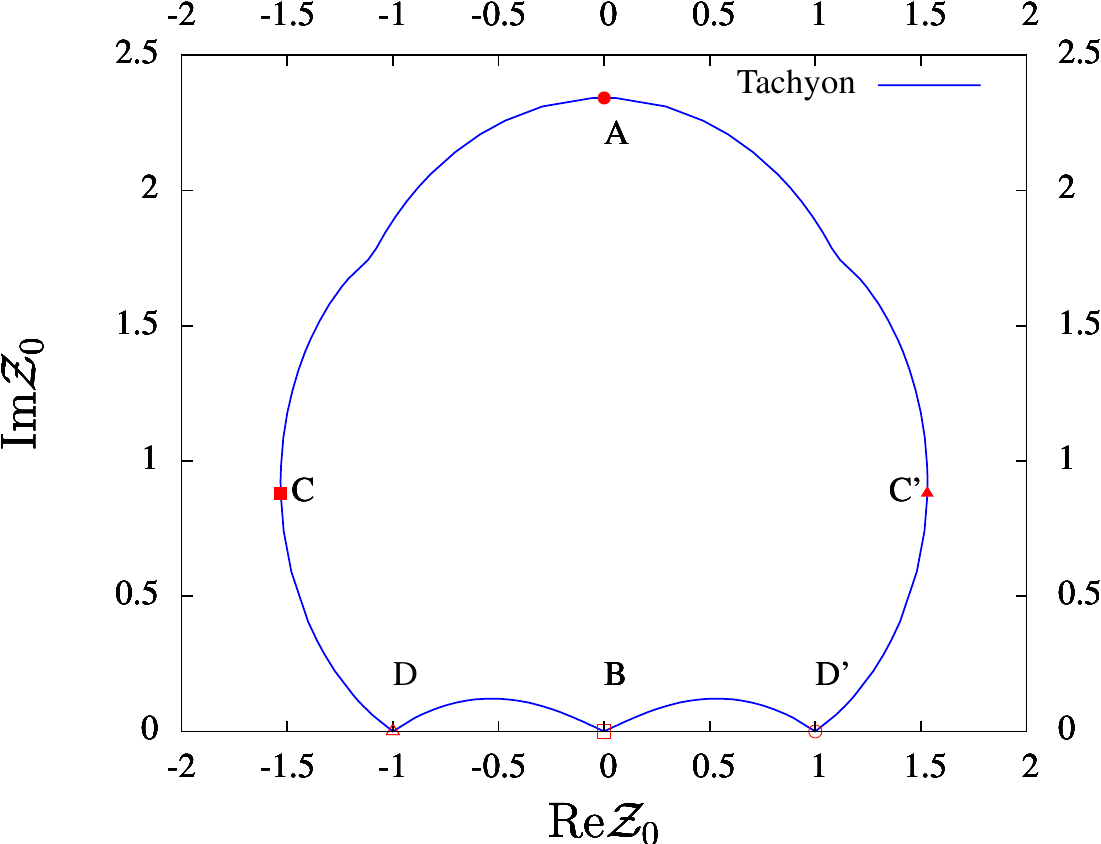}
\caption{Left: location of the Mkw solutions within the parameter space for the Supergravity models based on the $\,\mathfrak{so(4)}\,$ $B$-field reduction, highlighting the singular points. Right: set of VEVs of the modulus $\,\cZ\,$. Note that, up to discrete transformations, the Mkw extrema describe closed curves in both plots.}
\label{fig:SO4}
\end{figure}

\begin{itemize}
\item[$i)$] Points D and D' have an underlying $\,\fg=\mathfrak{iso(3)} \, + \, \mathfrak{so(4)}\,$ and are conjugate points with respect to the transformation (\ref{transWAll}). They are points of excluded Supergravity algebras \cite{deCarlos:2009fq}. As we flow towards these points, $\,\textrm{Im}\cS_{0} \rightarrow \infty\,$ while $\,\textrm{Im}\cT_{0} \, , \,  \textrm{Im}\cZ_{0} \rightarrow 0$. The tachyonic direction in field space is aligned with the Im$\cS$ modulus direction.

\item[$ii)$] The $\,\overline{\textrm{DD'}}\,$ line, going through the singular point B, has an underlying $\,\fg=\mathfrak{so(4)^{2}}\,$ Supergravity algebra. As we flow towards point B, the tachyon is still mostly aligned with Im$\cS\,$, and $\,\textrm{Im}\Phi_{0} \rightarrow 0\,$ for all the moduli fields, becoming once more a dynamical singularity in the moduli VEVs. However the axions  behave differently when approaching  the B point: $\,\textrm{Re}\cZ_{0} \rightarrow 0$, $\,\textrm{Re}\cS_{0} \rightarrow \pm \infty \,$ and $\,\textrm{Re}\cT_{0} \rightarrow \mp\infty\,$, with the upper sign choice if approaching from the left, and the other way around when approaching from the right. Notice, again, that this $\,\overline{\textrm{DD'}}\,$ line in the left plot of figure~\ref{fig:SO4} is smooth. 

\item[$iii)$] The $\,\overline{\textrm{DD'}}\,$ line going through the singular points C, C' and A, has an underlying $\,\fg=\mathfrak{so(3,1)} \, + \, \mathfrak{so(4)}\,$ Supergravity algebra. This path, shown in the left plot of figure~\ref{fig:SO4}, is discontinuous at points C, C' and A because of the vanishing of the Im$\cT\,$ modulus. The pairs of points with identical labels are conjugate points with respect to the transformation (\ref{transWT}). As we flow towards points C and C', $\,\textrm{Im}\cT_{0} \rightarrow 0\,$, and the tachyonic direction aligns $50\%$ in the Im$\cS$ direction and $50\%$ in the Re$\cS$ one. Finally, moving towards point A, $\,\textrm{Im}\cT_{0} \rightarrow 0\,$ and the tachyon is aligned with the Im$\cZ$ direction. These points C, C' and A are, then, dynamical singularities in the moduli VEVs.
\end{itemize}

\subsubsection*{The $\,\mathfrak{so(3,1)}\,$ models}
\label{sec:so31}

The last, but not least, Supergravity model based on a semisimple $B$-field reduction is $\,\mathfrak{so(3,1)}\,$. This model is defined in eqs~(\ref{kwModular}) by the superpotential
\beq
\begin{array}{ccl}
\mathcal{W}  & = &  |\Gamma|^{3/2} \left[ - 3 \, \mathcal{T} \, \left(\cz^3 + \cz\right) \,+ \, \mathcal{S} \, \left(   \eps_{2}  + 3 \,  \eps_{1} \, \cz  -  3  \, \eps_{2} \, \cz^{2}  -  \eps_{1} \, \cz^{3}  \right) - \right. \\ 
&-& \left. \xi_{3}  \left(  \eps_{1} \, - 3 \, \eps_{1} \, \cz^2  -  3  \, \eps_{2} \, \cz +  \eps_{2} \, \cZ^3  \right) + 3 \,  \xi_{7} \, \left(1 + \cz^2\right) \right] \ .
\end{array}
\label{WSO31}
\eeq
The most interesting feature of this model is that \textbf{it contains stable, Minkowski vacua} within a certain region of the parameter space as well as unstable Mkw solutions, like those of the previously analysed models, in a different one.
\begin{figure}[h!]
\centering
\includegraphics[width=8.5cm, height=6cm]{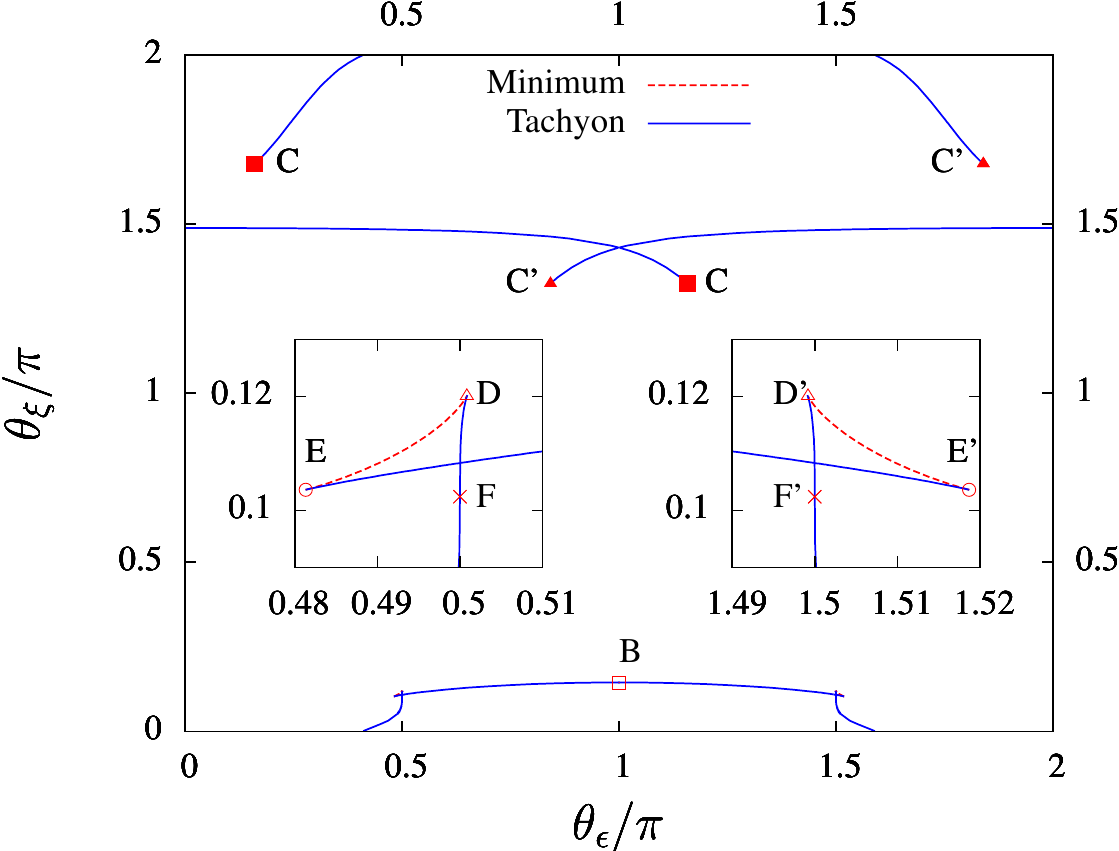}
\includegraphics[width=8.1cm, height=5.7cm]{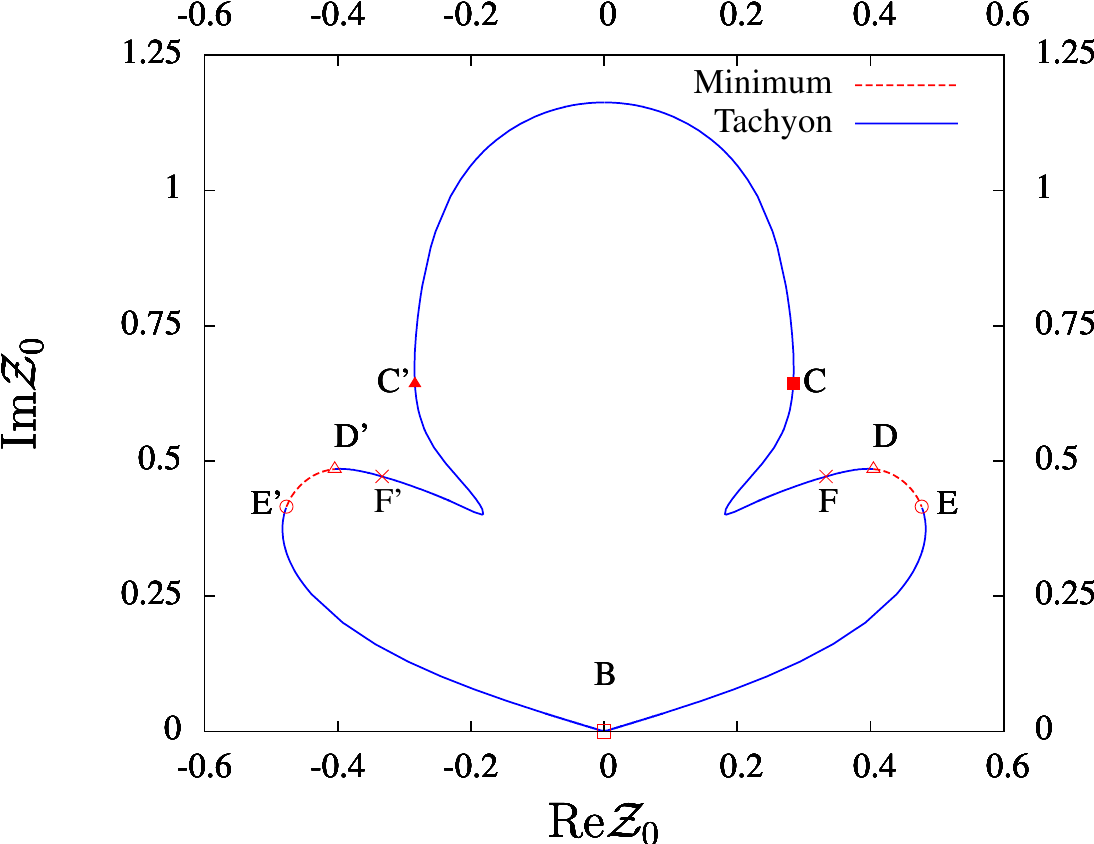}
\caption{Left: location of the Mkw solutions within the parameter space for the Supergravity models based on the $\,\mathfrak{so(3,1)}\,$ $B$-field reduction, highlighting the singular points. Right: set of VEVs of the modulus $\,\cZ\,$. Notice that, up to discrete transformations, the Mkw extrema describe closed curves in both plots.}
\label{fig:SO31}
\end{figure}
Another property of this model is that any point in the parameter space has a $\,\fg=\mathfrak{so(3,1)^{2}}\,$ Supergravity algebra underlying it. Therefore, any singularity in the moduli VEVs is a dynamical singularity. The entire set of Minkowski solutions are shown in figure~\ref{fig:SO31}.

With respect to the highlighted points in the figure, let us divide the parameter space in three pieces:  the $\,\overline{\textrm{DD'}}\,$ line going through the points C and C'; the $\,\overline{\textrm{EE'}}\,$ line going through the point B; and the $\,\overline{\textrm{DE}} \,\, \& \,\, \overline{\textrm{D'E'}}\,$ lines, containing the stable Mkw vacua:
\begin{itemize}
 \item[$i)$] At the points D, D', E and E', the Mkw extrema have a flat direction associated to volume directions~\footnote{At these points, the $\,2 \times 2\,$ reduced Hessian built from $\,\mathbb{V}(\cZ)\,$ in eq.~(\ref{v_eff}), becomes degenerate.}. This direction is, roughly, $\,58\%\,$ Im$\cS\,$ and $\,42\%\,$ Im$\cT\,$ at the D and D' points, whereas it becomes $\,72\%\,$ Im$\cS$, $\,25\%\,$ Im$\cT\,$ and $\,3\%\,$ Im$\cZ\,$  at the E and E' points. 

\item[$ii)$] The $\,\overline{\textrm{EE'}}\,$ line contains the singular point B. When moving towards it, the tachyon mostly aligns with the Im$\cS\,$ direction and $\,\textrm{Im}\Phi_{0} \rightarrow 0\,$ for all the moduli fields. The axions behave differently when approaching this point: $\,\textrm{Re}\cZ_{0} \rightarrow 0$, $\,\textrm{Re}\cS_{0} \rightarrow \mp \infty \,$ and $\,\textrm{Re}\cT_{0} \rightarrow \pm \infty\,$, with the upper sign choice if flowing from the left and the other choice when flowing from the right. Again, the $\,\overline{\textrm{EE'}}\,$ line in the left plot of figure~\ref{fig:SO31} is smooth.

\item[$iii)$] The $\,\overline{\textrm{DD'}}\,$ path goes through the singular points (F,F') and (C,C'). At (F,F')~\footnote{These F and F' points can be analytically computed and correspond to $\,(\theta_{\eps},\theta_{\xi})= \left( \pm \frac{\pi}{2} \,,\, \arctan\left(\frac{1}{3}\right)\right)$ together with the VEVs of $\,\cZ_{0}= \pm \frac{1}{3} + \frac{\sqrt{2}}{3} i\,$.}, it is discontinuous due to the double limits $\,\textrm{Im}\cS_{0} \rightarrow \dfrac{0}{0}\,$ and $\,\textrm{Im}\cT_{0} \rightarrow \dfrac{0}{0}\,$ in eq.~(\ref{min_x}). However, as we flow towards points C and C', a vanishing $\,\textrm{Im}\cS_{0} \rightarrow 0\,$ takes place, and the tachyonic direction mainly aligns with the Im$\cT\,$ volume direction. These points are, again, dynamical singularities in the moduli VEVs. Observe that points equally labeled in figure~\ref{fig:SO31} are conjugate points with respect to the transformation (\ref{transWS}).

\item[$iv)$] The $\,\overline{\textrm{DE}} \,\, \& \,\, \overline{\textrm{D'E'}}\,$ lines contain the stable Mkw vacua and will be explored separately.
\end{itemize}

There are two specially symmetric points which belong to part $iii)$ of the parameter space. The first one comes from noticing that this piece exhibits the novel feature of having a crossing at $\,(\theta_{\eps},\theta_{\xi})=(\pi , 1.43082 \pi )$. This crossing takes place in the parameter space, not in the moduli space, so two separate unstable Minkowski extrema 
\beq
\cZ_{0} = \pm 0.27527 + 0.80635 \, i  \hspace{2mm} , \hspace{2mm}  |\eps| |\xi|^{-1} \cS_{0} = \mp 0.87477 + 0.30709 \, i  \hspace{2mm} , \hspace{2mm}  |\xi|^{-1} \cT_{0} = \mp 0.44718 + 1.19429 \, i  \ , 
\label{2vacua}
\\[-1.5mm]
\eeq
with the tachyonic direction mostly along the Im$\cT\,$ volume direction, coexist at this point. The second point, located at $\,(\theta_{\eps},\theta_{\xi})=(0 , 1.48913 \pi)$, gives rise to an axion-vanishing unstable Mkw solution 
\beq
\cZ_{0} = 1.16280 \, i  \hspace{5mm} , \hspace{5mm}  |\eps| |\xi|^{-1}  \cS_{0} = 0.30849 \, i  \hspace{5mm} , \hspace{5mm}  |\xi|^{-1} \cT_{0} = 0.78019 \, i \ , 
\label{null_axions}
\eeq
invariant under the $\,\Phi \rightarrow - \Phi^{*}\,$ transformation of (\ref{transWAll}). The tachyonic direction is totally contained within the axion field space, with the relative contributions of $\,37\%\,$ for Re$\cS$, $\,40\%\,$ for Re$\cT\,$ and $\,23\%\,$ for Re$\cZ$. 
\\[6mm]
\textbf{$\overline{\textrm{DE}} \,\, \& \,\, \overline{\textrm{D'E'}}\,$ lines of stable vacua.} 
\\[1mm]
Let us look into the region within the parameter space that contains totally stable Minkowski vacua, namely, the $\,\overline{\textrm{DE}} \,\, \& \,\, \overline{\textrm{D'E'}}\,$ lines shown in figure~\ref{fig:SO31}.
\begin{figure}[h!]
 \centering
 \includegraphics[scale=0.75,keepaspectratio=true]{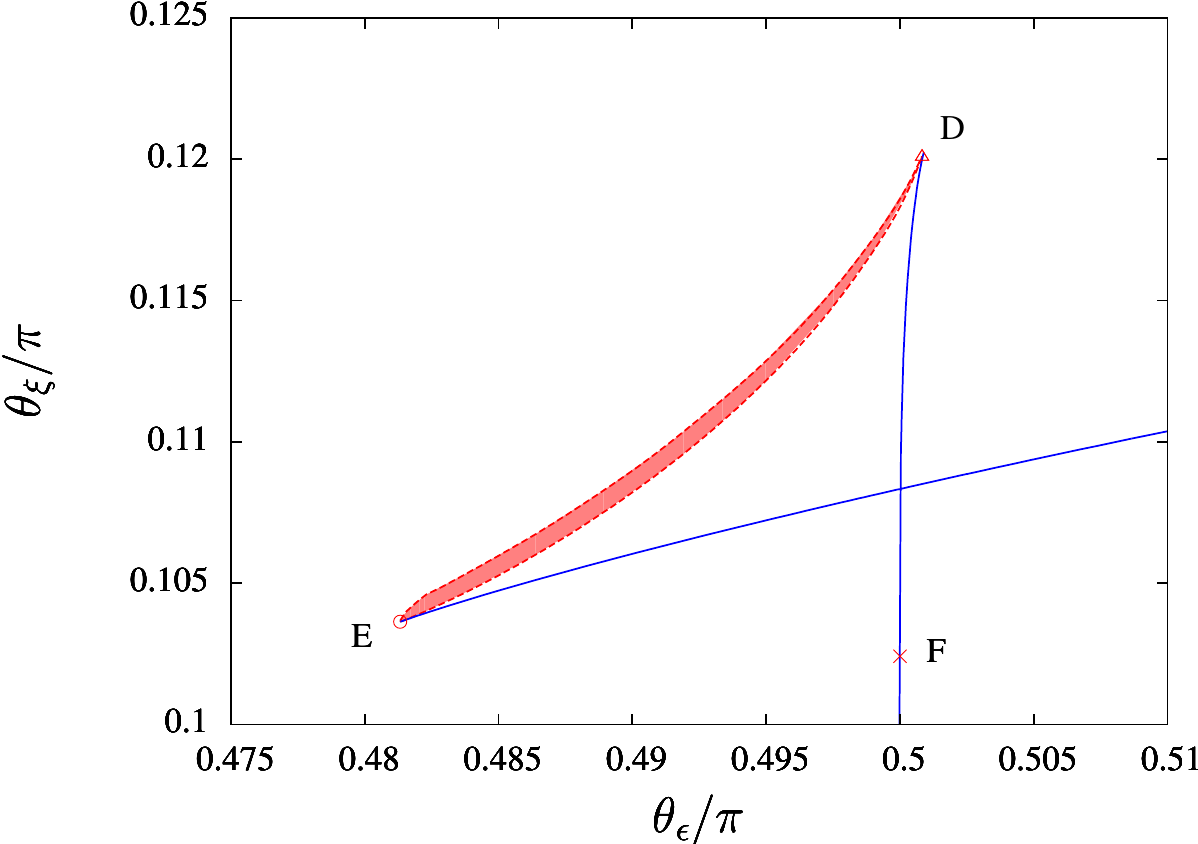}
 \caption{This figure shows the narrow band above the line of stable Mkw vacua containing stable, dS vacua.}
 \label{fig:dS_band}
\end{figure}
Provided a value for $\,\theta_{\eps}$ within the region $\overline{\textrm{DE}}\,$ (and equivalently for $\overline{\textrm{D'E'}}$), a stable dS vacuum emerges from varying the $\,\theta_{\xi}\,$ angle slightly with respect to its value at the Mkw vacuum, 
\beq
\theta_{\xi}^{\textrm{(dS)}} = \theta_{\xi}^{\textrm{(Mkw)}} + \delta\theta_{\xi}  \hspace{20mm} \textrm{ with } \,\,\hspace{10mm} \delta\theta_{\xi} > 0 \ .
\label{delta_theta_xi}
\eeq
There is a critical value, $\,\delta\theta_{\xi}^{*}$, beyond which the dS vacuum no longer exists\footnote{As an example, in the case of $\,\theta_{\epsilon}= \frac{49 \, \pi}{100}$, the moduli VEVs at the Mkw vacuum are given by $\,\cZ_{0} = 0.45089 + 0.46042 i \,$, $\,|\eps| |\xi|^{-1} \cS_{0} = -1.07734 + 1.28783 i \,$ and $\,|\xi|^{-1} \cT_{0} = 1.15629 + 0.60267 i $. This Mkw vacuum is compatible with $\,\theta_{\xi}^{\textrm{(Mkw)}}= 0.10821 \, \pi \,$, while the critical value for deforming it to dS is given by $\,\delta \theta_{\xi}^{*} = 0.00079 \, \pi$.}. This behaviour is represented in figure~\ref{fig:dS_band}. The dS vacua found in this way are deformations of the Mkw ones and are also stable along any direction in field space. Therefore, there is a narrow region above the line of Mkw vacua, shown in figure~\ref{fig:dS_band}, which incorporates dS stable vacua. Moreover if we choose $\,\delta\theta_{\xi} < 0$, the original Mkw vacuum becomes stable AdS. 

At these Mkw/dS vacua, Supersymmetry is broken by a non-vanishing F-term for all the moduli fields\footnote{In the case of $\,\theta_{\eps}=\frac{49 \, \pi}{100}$, the values of the F-terms at the Mkw vacuum are given by $\,|\Gamma|^{-\frac{3}{2}}\,|\xi|^{-1}\, F_{\cZ}= 4.00933 + 3.48324 i \,$, $\,|\Gamma|^{-\frac{3}{2}}\,|\eps|^{-1}\, F_{\cS}= 0.46460 - 0.00623 i \,$ and $\,|\Gamma|^{-\frac{3}{2}}\, F_{\cT}= -4.67506 + 5.76899 i $.}, i.e. $F_{\Phi=\cZ,\,\cS,\,\cT} \neq 0\,$. This agrees with the general results concerning the existence of non-supersymmetric, stable,  Minkowski vacua stated in refs~\cite{GomezReino:2006dk,GomezReino:2006wv}. Given that SUSY is broken by all directions considered here, seven complex ones in total, the constraint on the K\"ahler potential outlined in these works, formulated as the number of fields breaking SUSY being larger than
three, is fulfilled. Moreover, due to the F-term rescalings of (\ref{F-term_rescaling}), Supersymmetry breaking is dominated by $\,F_\cZ\,$ when $\,|\xi|\,$ increases, while it is dominated by $\,F_\cS\,$ as long as $\,|\eps|\,$ grows.

The (positive) smallest eigenvalue of the mass matrix is mostly associated to a combination of the Im$\cS$ and Im$\cT$ moduli fields, depending on the $\,|\eps|\,$ scaling parameter, i.e. it gets aligned with the Im$\cT$ volume when $\,|\eps|\,$ increases. At the Mkw vacua, the rest of the moduli masses are about a couple of order of magnitudes above the lightest one, unlike in scenarios including gaugino condensation or other non-perturbative effects \cite{Kachru:2003aw}. In the absence of large hierarchies we cannot split the stabilisation process into a $2+1$ fields problem, but the problem intrinsically becomes a $3$ fields one. This property is related to the fact that all the moduli are stabilised due to fluxes, so one would not expect to have mysterious cancellations in the mass terms in order to generate a hierarchy.

Finally, a dS saddle (tachyonic) point also appears close to these Mkw/dS stable vacua (in field space) with a much larger energy. This provides us with a natural scenario in which to investigate the possibilities for \textit{slow-roll} modular inflation to take place. We find the standard eta problem, i.e. $|\eta| \sim \mathcal{O}(10)$, of the inflationary models based on $\,\mathcal{N}=1$ Supergravity theories, when starting to roll from the dS saddle point to the Mkw/dS vacuum. This agrees with the results of \cite{Caviezel:2008tf,Flauger:2008ad} derived in the absence of non-geometric fluxes.

%
%

\section{Comparison with type IIA scenarios}
\label{sec:IIA}

The set of Supergravity models we have explored in the previous sections are dual to type IIA generalised flux models through applying three T-duality transformations along internal space directions \cite{STW,Aldazabal:2006up}. Several no-go theorems concerning the existence of Mkw/dS extrema in these type IIA generalised flux models have been stated as well as ways for circumventing them \cite{Hertzberg:2007wc,Silverstein:2007ac,Haque,Caviezel:2008tf,Flauger:2008ad}. In this section we will use the mapping introduced in ref.~\cite{deCarlos:2009fq} between the set of generalised flux models we derived in a type IIB with O3/O7-planes language, and their generalised type IIA dual flux models with O6-planes. Our purpose will be to investigate how the different sources of potential energy do conspire to produce the Mkw extrema we have found.   

In the type IIB side, the generalised set of NS-NS fluxes comprises the $\,\bar{H}_{3}\,$ and $\,Q\,$ fluxes. Going to the type IIA side, these fluxes map again to $\,\bar{H}_{3}\,$ and $\,Q\,$ flux components as well as new metric $\,\omega\,$ and non-geometric $\,R\,$ flux components. In the R-R flux sector, the situation looks similar. The $\,\bar{F}_{3}\,$ flux in the type IIB picture maps to components of the set $\,\bar{F}_{p}$, with $p=0,2,4$ and $6$, of type IIA fluxes \cite{STW,Aldazabal:2006up}. These $\,\bar{F}_{p}\,$ fluxes induce the set of IIA scalar potential contributions $\,V_{\bar{F}_{p}}$. The axions $\,\textrm{Re}\cS\,$ and $\,\textrm{Re}\cT$ enter the potential through the R-R piece $\,\VRR \subset \VIIA\,$, 
\beq
\VRR = \sum_{p=0\,(even)}^{6} V_{\bar{F}_{p}} = e^{\cK}\, \sum_{p=0\,(even)}^{6} \textrm{Im}\cZ^{(6-p)} \, \Big( f_{p}(\textrm{Re}\Phi)  \Big)^{2} \ ,
\label{Vaxions}
\eeq
where the functions $\,f_{p}\,$, with $p=0,2,4$ and $6$, depend on $\,\textrm{Re}\cS\,$ and $\,\textrm{Re}\cT$ linearly \cite{deCarlos:2009fq}.

The IIA/IIB correspondence between the contributions to the potential energy coming from localised sources results as follows. The O3-planes (D3-branes) in the type IIB models have to be interpreted as O6-planes (D6-branes) wrapping the $3$-cycle in the internal space which is invariant under the IIA orientifold action. In the following, we will refer to these O6/D6 sources as type 1. Finally, the O7-planes (D7-branes) in the type IIB side become O6-planes (D6-branes) wrapping a $3$-cycle invariant under the composition of the orbifold and the IIA orientifold actions \cite{Aldazabal:2006up,deCarlos:2009fq}. We will refer to these O6/D6 sources as type 2.

The scalar potential in the IIA dual Supergravity models then splits as
\beq
\VIIA=\VNS + \VRR + \Vloc \ ,
\label{VIIA}
\eeq
with $\,\VNS=V_{\bar{H}_{3}} + V_{\omega} + V_{Q} + V_{R}\,$ accounting for the generalised NS-NS fluxes, $\,\VRR=V_{\bar{F}_{0}} + V_{\bar{F}_{2}} + V_{\bar{F}_{4}} + V_{\bar{F}_{6}}\,$ accounting for the R-R fluxes and $\,\Vloc=\Vloc^{(1)}+\Vloc^{(2)}\,$ accounting for the (types 1 and 2) O6/D6 localised sources. In ref.~\cite{deCarlos:2009fq}, it was shown that the IIA duals of the IIB Supergravity models based on the $\,\mathfrak{nil}\,$ and $\,\mathfrak{iso(3)}\,$ $B$-field reductions yield $\,V_{Q}=V_{R}=0$, hence resulting in \textit{geometric} IIA flux models \cite{Villadoro:2005cu,Derendinger,Camara:2005dc,Grana:2006kf,Aldazabal:2007sn,Caviezel:2008ik}. This is also the case for the models based on the $\,\mathfrak{su(2)}+\mathfrak{u(1)^{3}}\,$ $B$-field reduction at the special circles $\,\theta_{\epsilon}=\pm \frac{\pi}{2}\,$ within the parameter space. Far from these circles as well as in those Supergravity models based on the $\,\mathfrak{so(4)}\,$ and $\,\mathfrak{so(3,1)}\,$ $B$-field reductions, $V_{Q}\neq0\,$ and/or $\,V_{R}\neq0$, giving rise to \textit{non-geometric} IIA flux models.

Let us now recall the most important results concerning the no-go theorems on the existence of dS/Mkw extrema in generalised IIA flux models mentioned above. For such solutions to exist, the terms in the scalar potential induced by the generalised NS-NS fluxes and the R-R fluxes have to satisfy
\beq
\begin{array}{rcrcrcr}
\left( V_{\omega} - V_{\bar{F}_{2}} \right) &+& 2 \,\left( V_{Q} - V_{\bar{F}_{4}} \right) &+& 3 \,\left( V_{R} - V_{\bar{F}_{6}} \right) &\ge& 0 \ , \\[2mm]
\left( V_{\bar{F}_{0}} - V_{\bar{H}_{3}} \right) &+& \left( V_{Q} - V_{\bar{F}_{4}} \right) &+& 2 \,\left( V_{R} - V_{\bar{F}_{6}} \right) &\ge& 0 \ ,
\label{no-go}
\end{array}
\eeq
where all the R-R flux-induced terms, $\,V_{\bar{F}_{p}}\,$, are positive definite, as well as the $\,V_{\bar{H}_{3}}\,$ and $\,V_{R}\,$ terms coming from the fluxes $\,\bar{H}_{3}\,$ and $\,R\,$, respectively\cite{deCarlos:2009fq}. The inequalities in (\ref{no-go}) are saturated at the Mkw extrema. Therefore, if restricting ourselves to the set of \textit{geometric} IIA flux models described above, there is a $\,V_{\bar{F}_{0}} \neq 0\,$ condition (non-vanishing Romans parameter) needed for having dS extrema. 

At this point, and before presenting our results in type IIA language, it is convenient to compare them with related work published in the literature. As we already mentioned in ref.~\cite{deCarlos:2009fq}, our framework is also that of ref.~\cite{Caviezel:2008tf}, which we have extended to include the set of generalised fluxes needed to restore T-duality.

In what concerns the several papers published on the existence of de Sitter solutions and no-go theorems, refs~\cite{Hertzberg:2007wc,Silverstein:2007ac,Haque}, there are some differences which are worth highlighting.  First of all, unlike refs~\cite{Silverstein:2007ac,Haque}, we do not consider KK five-branes. Neither we consider NS5-branes as they do in refs~\cite{Hertzberg:2007wc,Haque}. However the most important difference with all these works is the fact that our minimisation procedure considers the dependence of the potential on the axions which are treated as dynamical variables. While in the references pointed out they are set to constant values and do not feature in the scalar potential.

There are also substantial differences between our work with that of ref.~\cite{Flauger:2008ad}. On the one hand, these authors consider K\"ahler and complex structure moduli in addition to the dilaton and volume moduli considered in the previous works reviewed here. However, the potential contains the effect of just geometric fluxes, in addition to the usual NS-NS 3-form flux, R-R fluxes and O6/D6 sources. Nevertheless they manage to find a couple of $\,\mathbb{Z}_2 \times \mathbb{Z}_2\,$ orbifold models that, within their working numerical precision, are compatible with de Sitter vacua. These are both anisotropic models and cannot, therefore, be compared to ours. In any case it is worth mentioning that, throughout their analysis, these authors find plenty of solutions with one tachyonic direction, just as it happens in our analysis.

\subsection{Minkowski extrema in \textit{geometric} type IIA flux models}

As we have stated above, there are three sets of type IIB Supergravity models that become dual to \textit{geometric} type IIA flux models with
\beq
V_{Q}=V_{R}=0 \ .
\label{VQVRnull}
\eeq
They are the models based on the $\,\mathfrak{nil}\,$ and $\,\mathfrak{iso(3)}\,$ $B$-field reductions together with those based on the $\,\mathfrak{su(2)}+\mathfrak{u(1)^{3}}\,$ reduction at the circles defined by $\,\theta_{\epsilon}=\pm \frac{\pi}{2}\,$ in the parameter space.

\begin{figure}[h!]
\centering
\includegraphics[scale=0.75,keepaspectratio=true]{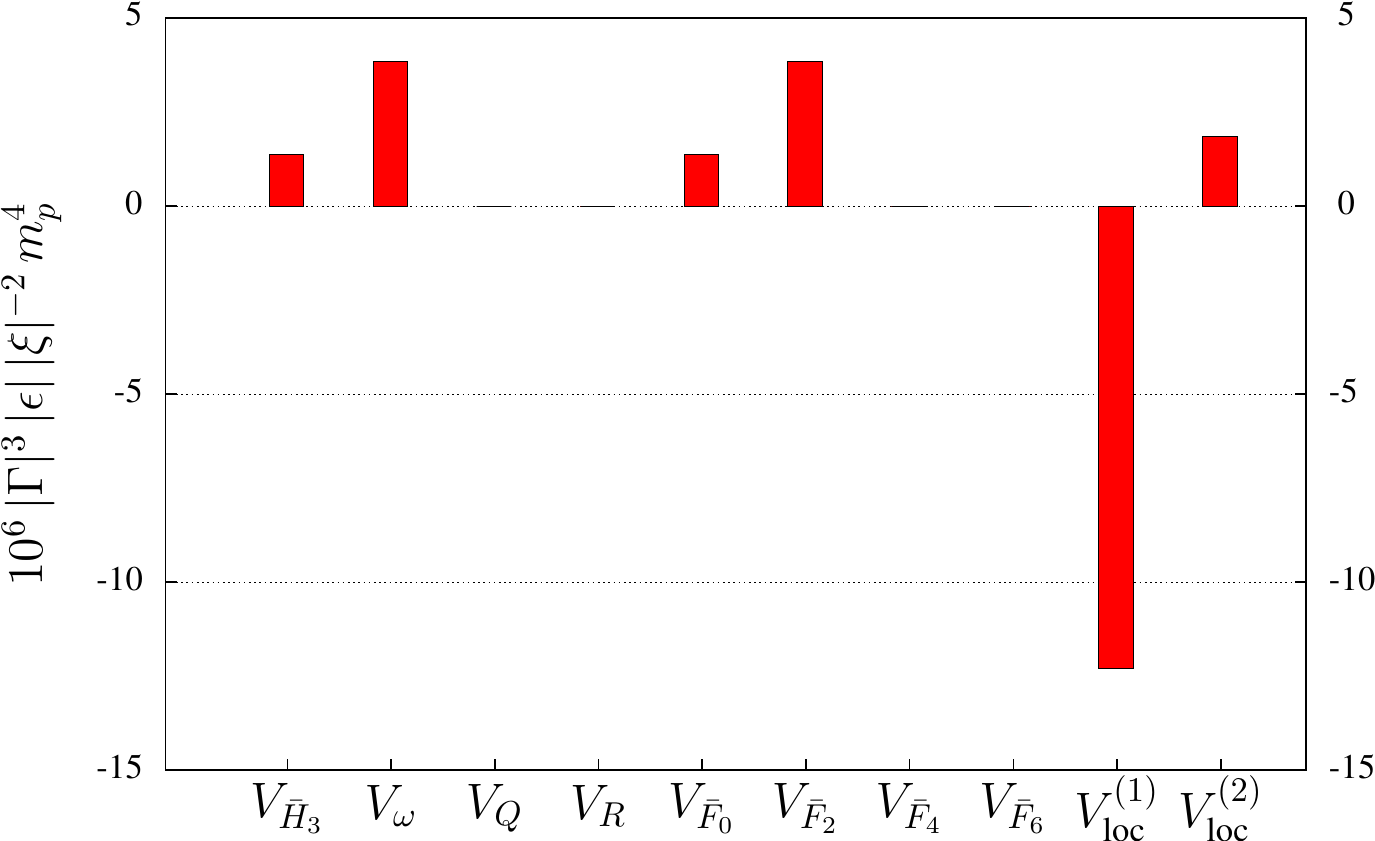}
\caption{IIA dual contributions to the scalar potential at the Mkw extrema for the Supergravity models based on the $\,\mathfrak{iso(3)}\,$ $B$-field reduction. They are computed at the circle $\,\theta_{\epsilon}=\frac{3 \pi}{4}\,$ in the parameter space which implies $\,\theta_{\xi} = 0.22375 \, \pi\,$ and the moduli VEVs of $\,\cZ_{0}= 0.30920 + 0.11495  \, i \,$, $\,|\eps| |\xi|^{-1} \cS_{0} = -0.00171 + 0.01276 i\,$ and $\,|\xi|^{-1} \cT_{0} = 0.01579 + 0.00092 i$. It can be seen that $\,V_{Q}=V_{R}=V_{\bar{F}_{4}}=V_{\bar{F}_{6}}=0\,$ as well as $\,V_{\bar{H}_{3}} = V_{\bar{F}_{0}}>0\,$ and $\,V_{\omega} = V_{\bar{F}_{2}}>0$.}
\label{fig:Hist_iso}
\end{figure}

A common feature in all these IIA dual \textit{geometric} models is that only the $\,f_{4}\,$ and $\,f_{6}\,$ functions appearing in (\ref{Vaxions}) depend (linearly) on the Re$\cS$ and Re$\cT$ axions. Then, their stabilisation conditions, provided $\,\textrm{Im}\cZ_{0} \neq 0$, translate into
\beq
V_{\bar{F}_{4}}=V_{\bar{F}_{6}}=0 \ .
\label{VF4VF6null}
\eeq
Substituting (\ref{VQVRnull}) and (\ref{VF4VF6null}) into the inequalities of (\ref{no-go}), we obtain, for any Minkowski extremum, that
\beq
V_{\bar{H}_{3}} = V_{\bar{F}_{0}}  \hspace{1cm} \textrm{and} \hspace{1cm}  V_{\omega} = V_{\bar{F}_{2}} \ ,
\eeq
so $\,\VNS = \VRR >0\,$ at such extrema\footnote{Notice that due to the positiveness of $\,V_{\bar{F}_{2}}$, the $\,V_{\omega}\,$ contribution to the scalar potential coming from the (negative) curvature of the internal space (induced by the metric flux $\,\omega\,$) results also positive as it was stated in \cite{Silverstein:2007ac,Haque}.}. Then, the negative energy contribution needed to set $\,\VIIA=0\,$ in (\ref{VIIA}) will come from the localized sources, i.e. $\,\Vloc<0\,$ (see figure~\ref{fig:Hist_iso}).

The IIB Supergravity models based on the $\,\mathfrak{nil}\,$ reduction were found in section \ref{sec:Nil} not to accommodate for Mkw extrema while those based on the $\,\mathfrak{su(2)}+\mathfrak{u(1)^{3}}\,$ reduction at the circles $\,\theta_{\epsilon}=\pm \frac{\pi}{2}\,$ were excluded to possess Mkw extrema in ref.~\cite{deCarlos:2009fq}. Therefore, the Mkw extrema we found in the Supergravity models based on the $\mathfrak{iso(3)}$ $B$-field reduction, constitute the entire set of \textit{geometric} IIA dual Minkowski flux extrema for the isotropic $\,\mathbb{Z}_{2} \times \mathbb{Z}_{2}\,$ orbifold. These extrema have an underlying $\,\fg=\mathfrak{so(4)} \,\oplus_{{\mathbb{Z}_{3}}}\, \mathfrak{u(1)^{6}}\,$ Supergravity algebra \cite{deCarlos:2009fq}.

The IIA dual contributions to the scalar potential at the \textit{geometric} Mkw flux extrema are shown in figure~\ref{fig:Hist_iso} (where $m_{p}=1/\sqrt{8 \pi G} \approx 2 \times 10^{18} \, \textrm{GeV}$). Although they are plotted for a particular point within the parameter space, the profile of the contributions does not change when moving from one point to another, due to the scaling properties explained in section~\ref{sec:Numerical}. Observe that the negative energy contribution needed to obtain $\,\VIIA=0\,$ comes from the O6/D6 sources wrapping the 3-cycle invariant under the orientifold action (type 1). Specifically, from O6-planes which carry negative charge. Moreover, additional positive energy coming from type 2 D6-branes with positive charge is also required. These type 2 sources are forbidden in the $\,\mathbb{Z}_{2}\,$ orbifold compactifications of \cite{STW,Aldazabal:2006up}, so these \textit{geometric} IIA dual Mkw extrema are not expected to exist there.

Finally, for these IIB Supergravity models based on the $\mathfrak{iso(3)}$ $B$-field reduction, the IIA dual Romans parameter which generates the $V_{\bar{F}_{0}}$ contribution required for having dS extrema, reads
\beq
f_{0}^{2} = 4 \, |\Gamma|^{3} \, |\epsilon|^{2} \, |\xi|^{2} \, (\sin\theta_{\eps})^{2} \,(\cos\theta_{\xi})^{2} \ ,
\label{Romans_iso} 
\eeq
so it vanishes at the A, A' and B singular points shown in figure~\ref{fig:iso}. Far from these points, an unstable dS extremum emerges from varying the $\,\theta_{\xi}\,$ angle slightly with respect to its value at the Mkw extremum, $\,\theta_{\xi}^{\textrm{(dS)}} = \theta_{\xi}^{\textrm{(Mkw)}} + \delta\theta_{\xi}\,$ with $\,\delta\theta_{\xi} > 0$, as it has been previously explained for the case of the stable dS vacua in the $\mathfrak{so(3,1)}$-based models. Also a critical value $\,\delta\theta_{\xi}^{*}\,$ appears beyond which dS solutions no longer exist.

\subsection{Minkowski extrema in \textit{non-geometric} type IIA flux models}

Now we present the IIA dual energy contributions at the Mkw extrema for the Supergravity models which are \textit{non-geometric} type IIA generalised flux models. These models are those based on the $\,\mathfrak{su(2)} + \mathfrak{u(1)^3}\,$ (with $\,\theta_{\eps} \neq \pm \frac{\pi}{2}\,$), $\,\mathfrak{so(4)}\,$ and $\,\mathfrak{so(3,1)}\,$ $B$-field reductions.

\subsubsection*{The $\,\mathfrak{su(2)} + \mathfrak{u(1)^3}\,$ models}

\begin{figure}[h!]
\centering
\includegraphics[scale=0.75,keepaspectratio=true]{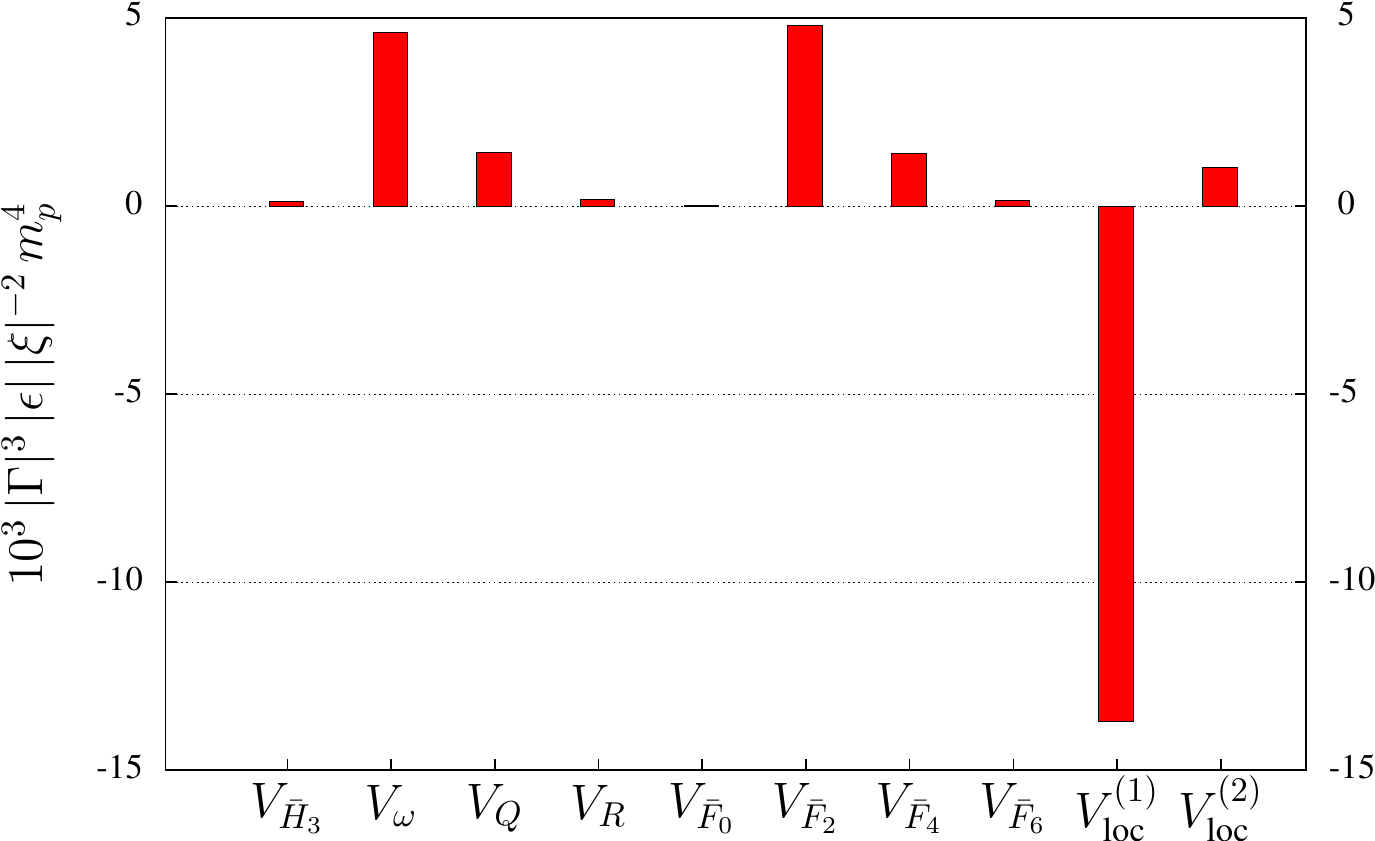}
\caption{IIA dual contributions to the scalar potential at the Mkw extrema for the Supergravity models based on the $\,\mathfrak{su(2)}+\mathfrak{u(1)^{3}}\,$ $B$-field reduction. They are computed, again, at the circle $\,\theta_{\epsilon}=\frac{3 \pi}{4}\,$ in the parameter space implying this time $\,\theta_{\xi} = 0.13055 \, \pi\,$ and the moduli VEVs of $\,\cZ_{0} = 0.99368 + 0.55061  \, i\,$, $\,|\eps| |\xi|^{-1} \cS_{0} = -1.01524 + 0.28041 i\,$ and $\,|\xi|^{-1} \cT_{0} = 0.82169 + 0.01611 i$.}
\label{fig:Hist_direct}
\end{figure}

As it was stated in section~\ref{sec:non-semisimple_models}, these models also have the scaling property of the \textit{geometric} IIA dual models. Therefore, their profile of energy contributions, shown in figure~\ref{fig:Hist_direct}, does not change from one point within the parameter space to another. 

These \textit{non-geometric} type IIA dual flux models (note that $\,V_{Q} \neq 0\,$ and $\,V_{R} \neq 0\,$) need again of localised sources to achieve Minkowski (unstable) solutions. Analogously to the \textit{geometric} case, type 1 O6-planes and type 2 D6-branes are required, as it can be seen in figure~\ref{fig:Hist_direct}. Also each contribution in $\VNS$ and $\VRR$ is positive at the Mkw solutions. Finally, unstable dS solutions can again be obtained by deforming these Mkw extrema, as for the \textit{geometric} IIA dual models.

\subsubsection*{The $\,\mathfrak{so(4)}\,$ models}

The next Supergravity models whose IIA duals become \textit{non-geometric} flux models are those based on the semisimple $\,\mathfrak{so(4)}\,$ $B$-field reduction. The contributions to the potential energy at the Minkowski extrema do not fit a unique pattern, as it has been the case for the Supergravity models analysed so far. Such contributions do depend on the point in the parameter space under consideration, since the scaling property (\ref{Z_scaling}) is no longer present in these models. 
\begin{figure}[h!]
\centering
\includegraphics[scale=0.6,keepaspectratio=true]{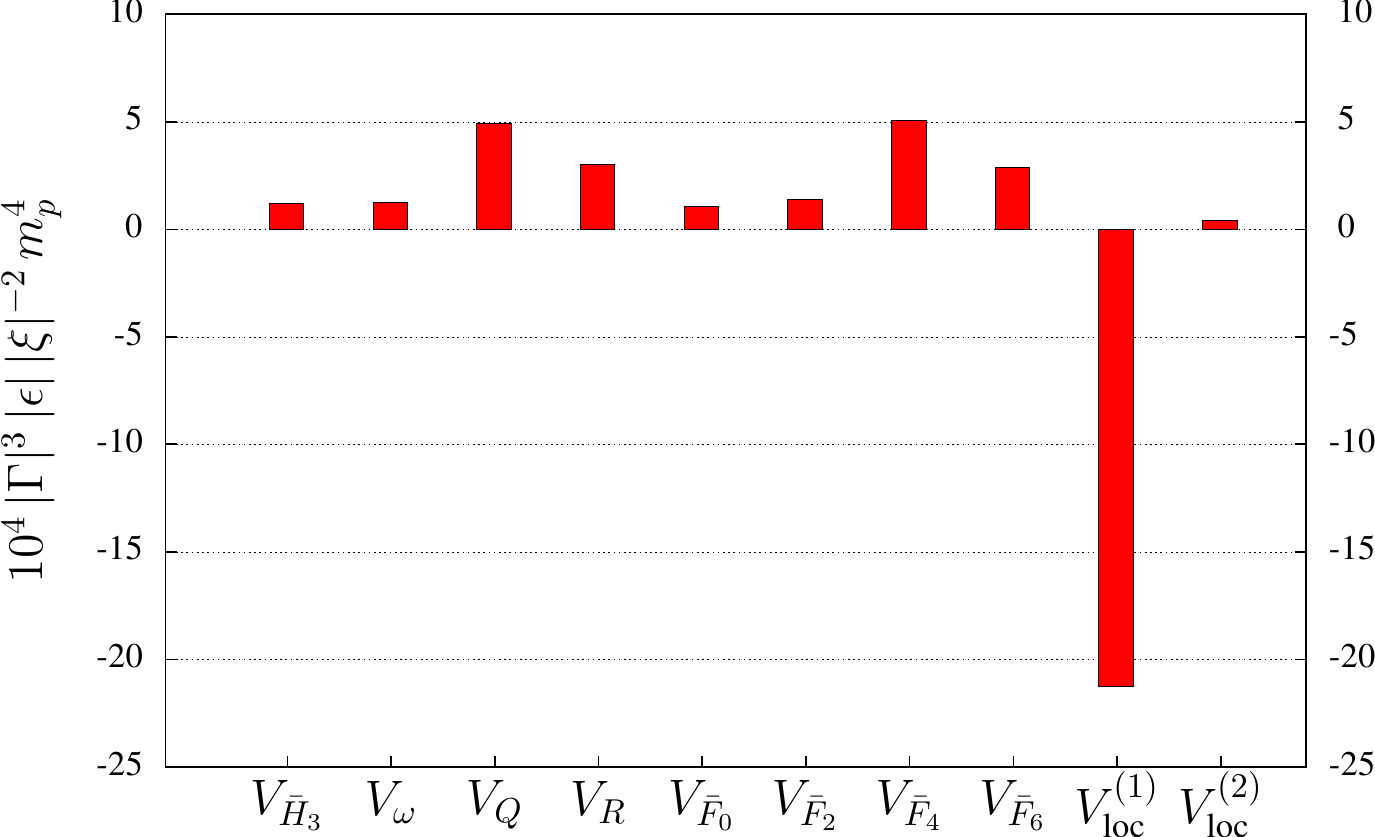}
\includegraphics[scale=0.6,keepaspectratio=true]{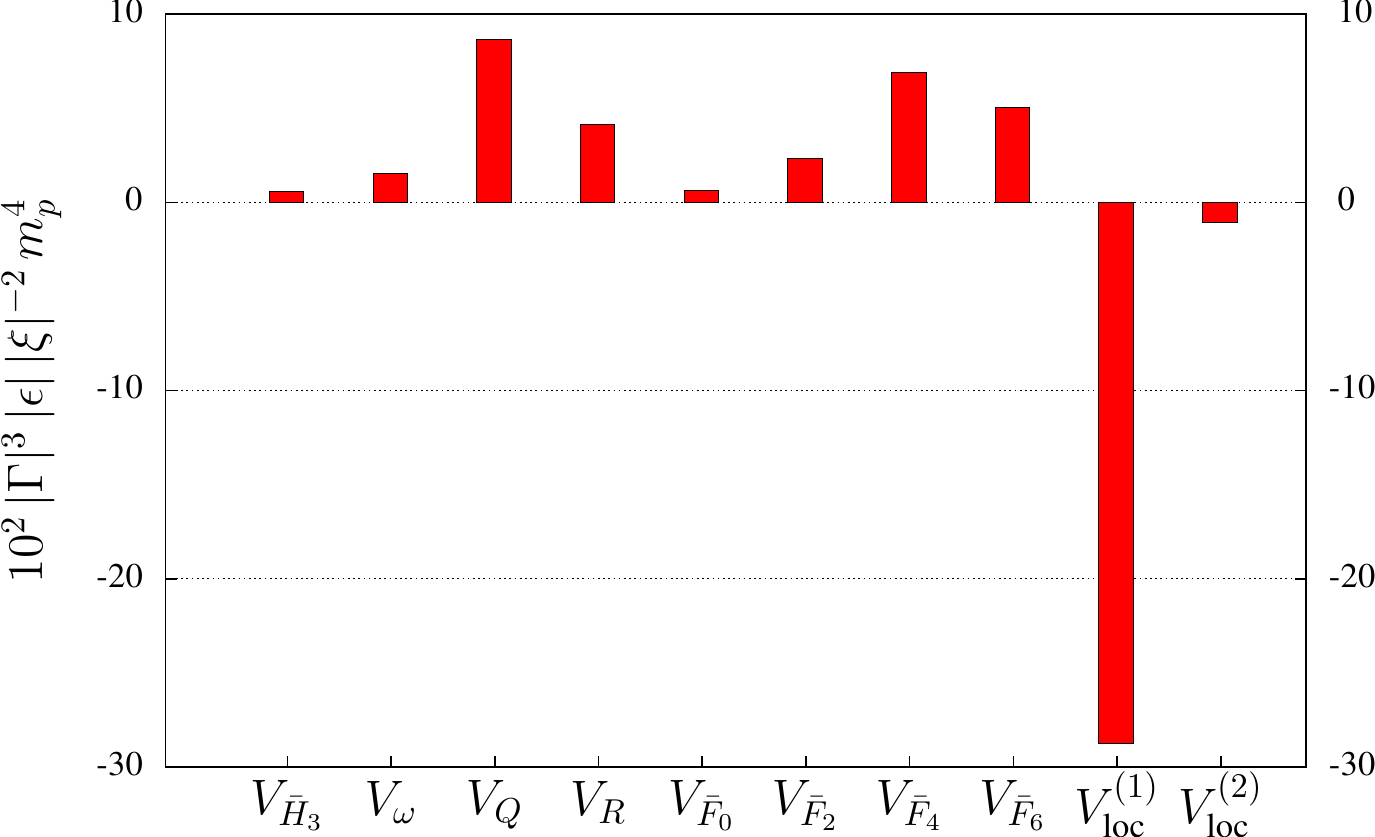}
\caption{IIA dual contributions to the scalar potential at a Mkw extremum for the Supergravity models based on the $\,\mathfrak{so(4)}\,$ $B$-field reduction. In the left plot, they are computed at the circle $\,\theta_{\epsilon}=  \frac{1255}{1000}\pi\,$ which belongs to the $\,\overline{\textrm{DC}}\,$ piece of the parameter space and implies $\,\theta_{\xi} = 0.40225 \, \pi\,$ together with the moduli VEVs of $\,\cZ_{0} = -1.48023 + 0.59103 i\,$, $\,|\eps| |\xi|^{-1} \cS_{0} = -1.15549 + 3.44203 i\,$ and $\,|\xi|^{-1} \cT_{0} = 0.13871 + 0.00708 i$. In the right plot they are computed at the circle $\,\theta_{\epsilon}=\frac{3 \pi}{8}\,$ which belongs to the $\,\overline{\textrm{CA}}\,$ piece of the parameter space and implies $\,\theta_{\xi} = 1.93558 \, \pi\,$ together with the moduli VEVs of $\,\cZ_{0} = -0.73422 + 2.12313 i\,$, $\,|\eps| |\xi|^{-1} \cS_{0} = -0.57723 + 0.74810 i\,$ and $\,|\xi|^{-1} \cT_{0} = 0.33682 + 0.04399 i$.}
\label{fig:Hist_so4_localised}
\end{figure}

In order to illustrate the above statement, let us recall the form of the contributions to the scalar potential coming from the localised sources. They were computed in ref.~\cite{deCarlos:2009fq}, and given by  
\beq
\Vloc^{(1)}= - \frac{|\epsilon|\, |\Gamma|^{3}}{4\, |\xi|^{2} \, \textrm{Im}\cT^{3}} \,\cos\theta_{\xi} \hspace{10mm} \textrm{and} \hspace{10mm} \Vloc^{(2)}= \frac{3\,|\epsilon|\, |\Gamma|^{3}}{4\, |\xi|^{2} \, \textrm{Im}\cT^{2} \, \textrm{Im}\cS} \, \sin\theta_{\xi} \ ,
\label{V_sources_semi}
\eeq
for the Supergravity models based on semisimple $B$-field reductions. Then $\,\Vloc^{(1)}=0\,$ at the D and D' singular points shown in figure~\ref{fig:SO4}, while $\,\Vloc^{(1)}<0\,$ in all the Mkw solutions. On the other side, $\,\Vloc^{(2)}=0\,$ at the singular point A, whereas $\,\Vloc^{(2)}>0\,$ for the Mkw solutions along the $\,\overline{\textrm{CC'}}\,$ line that goes through point B, and  $\,\Vloc^{(2)}<0\,$ if doing so through point A. An example is shown in figure~\ref{fig:Hist_so4_localised}, where the sign of the energy contribution provided by type 2 localised sources is different for the two Mkw solutions. In the left plot type 2 D6-branes are required, while type 2 O6-planes are needed in the right one.

Finally, one observes that the flux-induced $\,\cP_{2,3}(\cZ)\,$ polynomials for these models reduce to those of the \textit{geometric} ($\,\mathfrak{iso(3)}$-based) models around $\,\cZ =0$, as it can be seen from their form in table~\ref{tablecP23}. As long as we take the limit $\,\theta_{\eps} \rightarrow \pi\,$, the profile (up to some scale factor) of the energy contributions to the Mkw extrema tend to that of the \textit{geometric} models in figure~\ref{fig:Hist_iso}.
Once more, unstable dS extrema can be obtained by a continuous deformation of the Mkw solutions, namely, by taking $\,\theta_{\xi} \rightarrow \theta_{\xi} + \delta \theta_{\xi}\,$ for a given $\,\theta_{\epsilon}\,$ circle.

\subsubsection*{The $\,\mathfrak{so(3,1)}\,$ models}

Let us conclude by looking into the energy contributions to the Mkw extrema for the IIA duals of the Supergravities models based on the $\,\mathfrak{so(3,1)}\,$ $B$-field reduction. As for the previous semisimple models, such contributions depend critically on the specific point within the parameter space under consideration.
\begin{figure}[h!]
\centering
\includegraphics[scale=0.75,keepaspectratio=true]{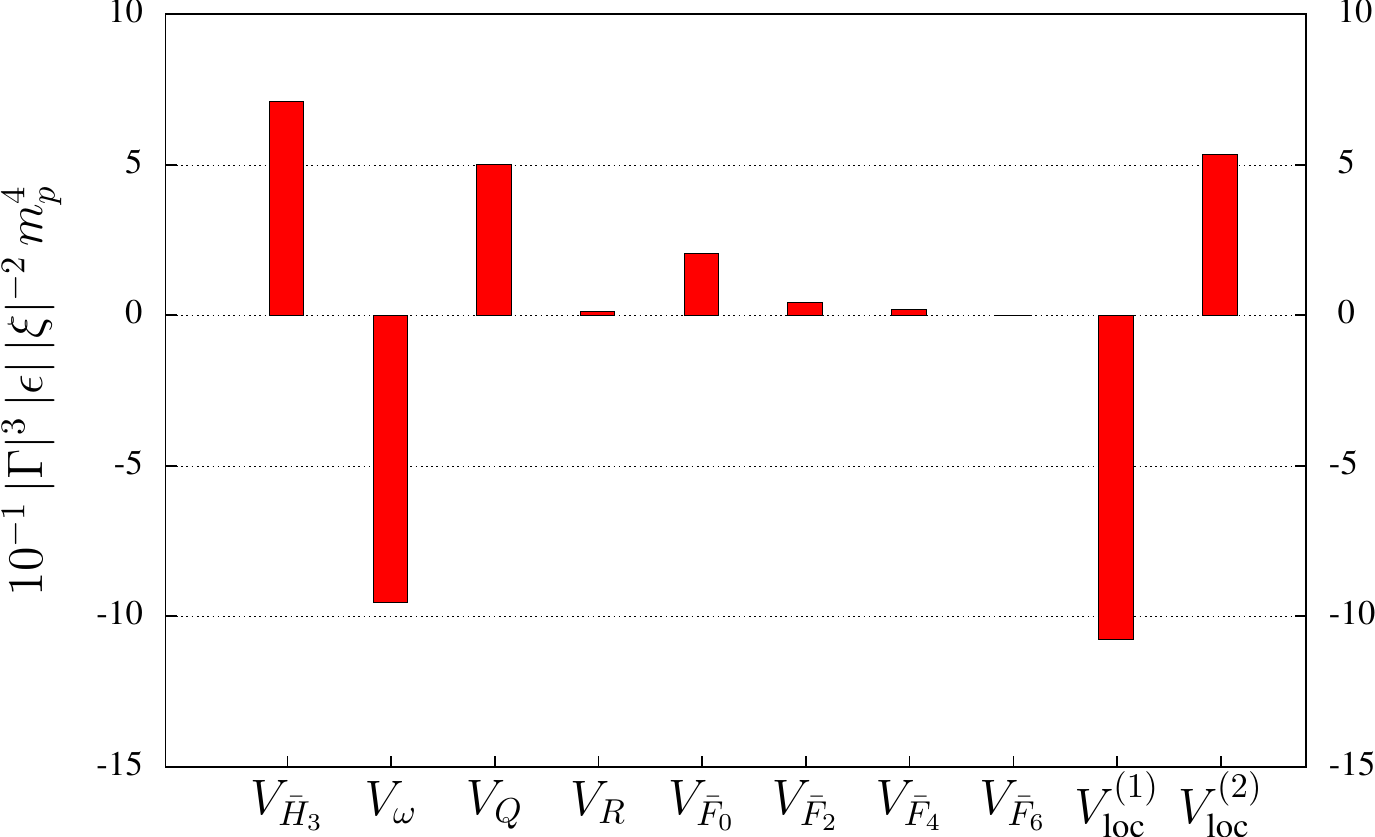}
\caption{IIA dual contributions to the scalar potential at a stable Mkw vacuum for the Supergravity models based on the $\,\mathfrak{so(3,1)}\,$ $B$-field reduction. They are computed at the circle $\,\theta_{\epsilon}= \frac{49 \pi}{100} \,$ which belongs to the $\,\overline{\textrm{DE}}\,$ piece of the parameter space and implies $\,\theta_{\xi} = 0.10821 \, \pi\,$, together with the moduli VEVs of $\,\cZ_{0} = 0.45089 + 0.46042 i \,$, $\,|\eps| |\xi|^{-1} \cS_{0} = -1.07734 + 1.28783 i \,$ and $\,|\xi|^{-1} \cT_{0} = 1.15629 + 0.60267 i $.}
\label{fig:Hist_so31_stable}
\end{figure}

The set of Minkowski solutions for this model is shown in figure~\ref{fig:SO31}, where a narrow region within the parameter space, that of the $\,\overline{\textrm{DE}} \,\, \& \,\, \overline{\textrm{D'E'}}\,$ lines, was found to contain \textbf{stable vacua}. At these stable vacua, $\,V_{\omega}<0\,$ and $\,\Vloc^{(1)}<0\,$, while the rest of the contributions to the scalar potential are positive. Then, these stable vacua need  type 1 O6-planes and type 2 D6-branes to exist. As long as we flow between the points D and E in figure~\ref{fig:SO31}, the main contributions to $\,\frac{|\xi|^{2}}{|\Gamma|^{3}\,|\eps|\,m_{p}^{4}}\, \VIIA\,$ change from being of order $\,\mathcal{O}(10^{-2})\,$ around the point D, to become of order $\,\mathcal{O}(1)\,$ around the point E. An intermediate point in the $\,\overline{\textrm{DE}}\,$ line is shown in figure~\ref{fig:Hist_so31_stable}.

\begin{figure}[h!]
\centering
\includegraphics[scale=0.75,keepaspectratio=true]{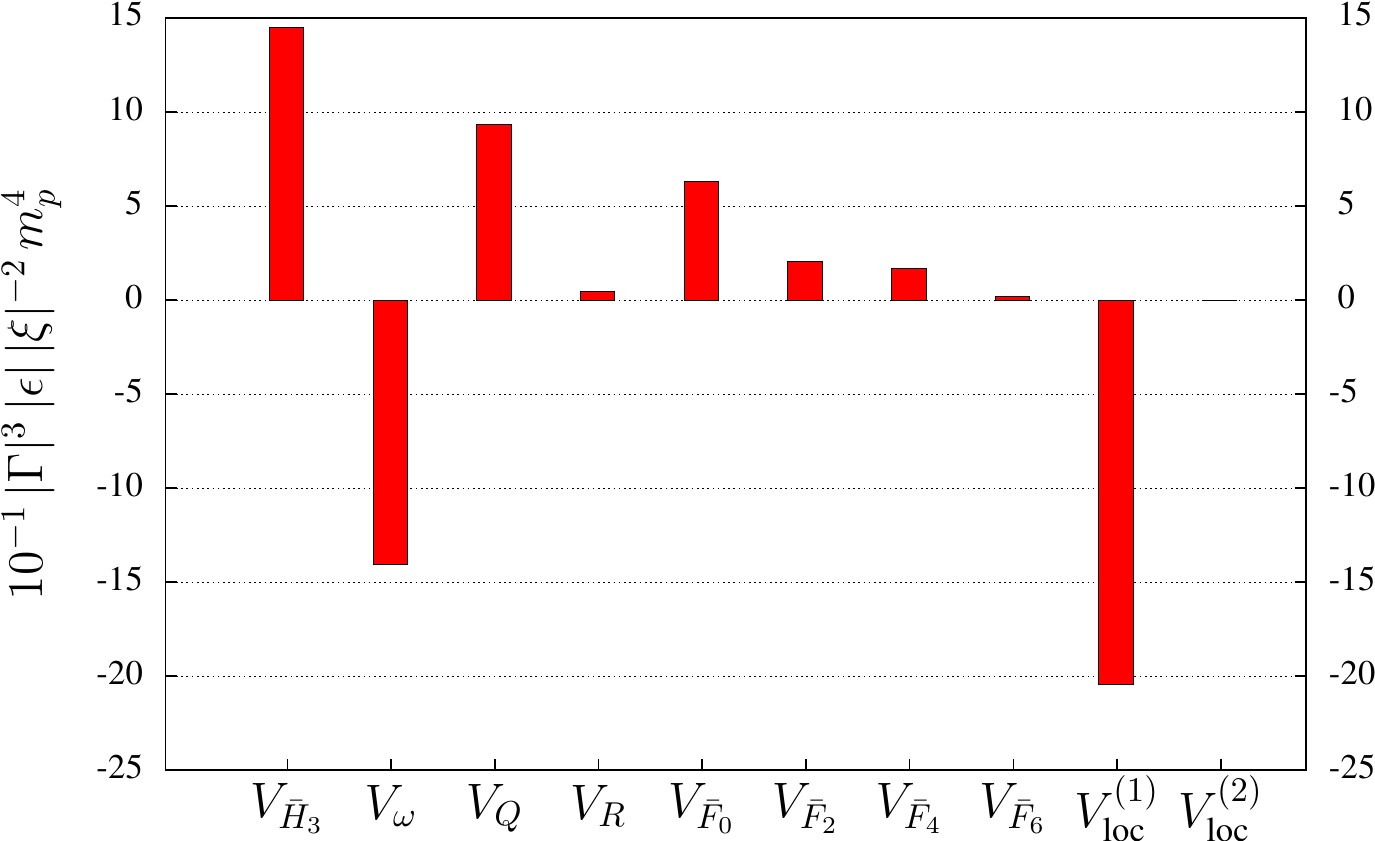}
\caption{IIA dual contributions to the scalar potential at a unstable Mkw solution for the Supergravity models based on the $\,\mathfrak{so(3,1)}\,$ $B$-field reduction. It is computed at the point $\,(\theta_{\eps} , \theta_{\xi}) = ( 0.40904 \pi , 0)\,$ which belongs to the $\,\overline{\textrm{CD}}\,$ line in the parameter space. The VEVs for the moduli fields result in $\,\cZ_{0} = 0.18657 + 0.41905 i \,$, $\,|\eps| |\xi|^{-1} \cS_{0} = 0.12569 + 0.32326 i \,$ and $\,|\xi|^{-1} \cT_{0} = 0.76855 + 0.49656 i $.}
\label{fig:Hist_so31_V2null}
\end{figure}

For these Supergravity models, the contributions to the potential energy coming from localised sources are still given by (\ref{V_sources_semi}). By inspection of figure~\ref{fig:SO31}, we conclude that there are unstable Mkw solutions having $\,\Vloc^{(2)} \gtrless 0$. Even more, there is a particularly interesting solution with $\,\Vloc^{(2)} = 0$. It is located at the point $\,(\theta_{\eps} , \theta_{\xi}) = ( 0.40904 \pi , 0)\,$ within the parameter space, and its profile of the contributions to $\,\VIIA\,$ is shown in figure~\ref{fig:Hist_so31_V2null}.
Naturally, its image point under the transformation $\,\Phi \rightarrow - \Phi^{*}\,$ of (\ref{transWAll}) is also a solution with $\,\Vloc^{(2)} = 0$. These unstable Mkw solutions are the only ones that would also exist in the $\,\mathbb{Z}_{2}\,$ orbifold compactification of refs~\cite{STW,Aldazabal:2006up}, that does not allow type 2 O6/D6 sources. In the absence of such sources, these unstable solutions could presumably be lifted to solutions of a $\,\mathcal{N}=4\,$ gauged Supergravity \cite{Guarino:2008ik,Derendinger,Aldazabal:2008zza,Roest:2009dq,Dall'Agata:2009gv,Avramis:2009xi} built from an electric-magnetic gauging \footnote{We thank G.~Dibitetto and D.~Roest for discussions on this point.}. 

\begin{figure}[h!]
\centering
\includegraphics[scale=0.60,keepaspectratio=true]{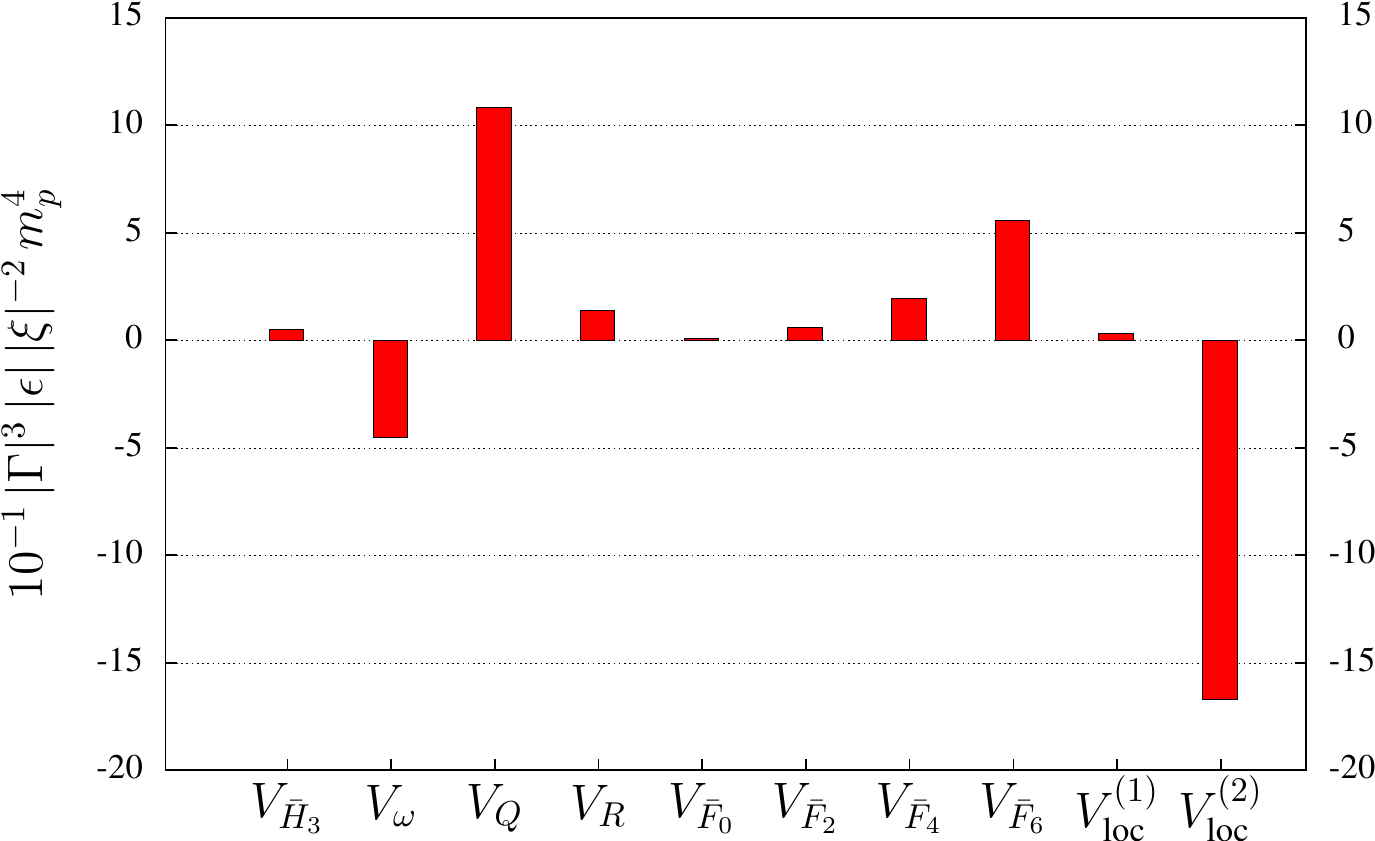}
\includegraphics[scale=0.60,keepaspectratio=true]{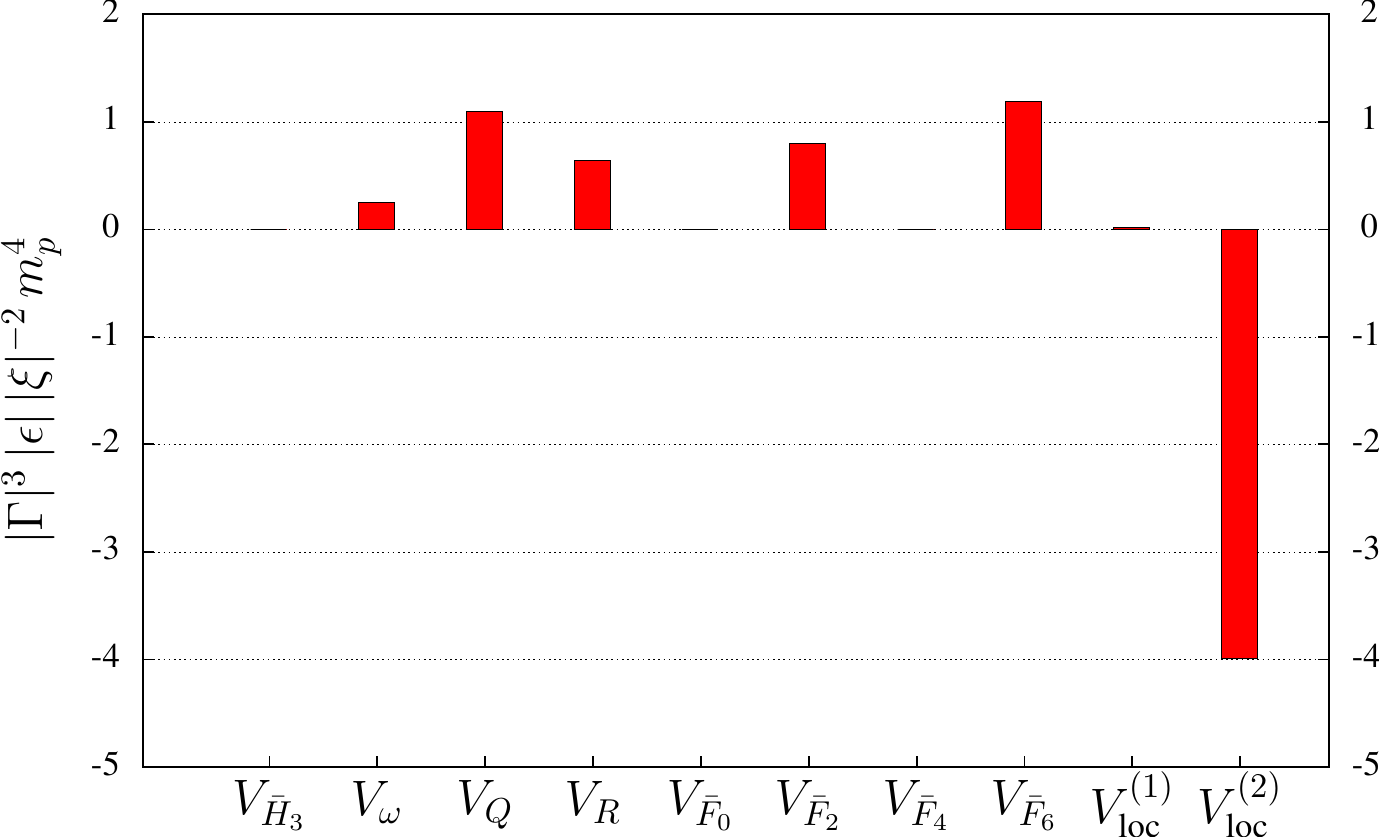}
\caption{Left: IIA dual contributions to the scalar potential at the unstable Mkw solutions that coexist at the point $\,(\theta_{\eps},\theta_{\xi})=(\pi , 1.43082 \pi )$ in the parameter space. Right: same for the axion-vanishing solution at the point of parameter space given by $\,(\theta_{\eps},\theta_{\xi})=(0 , 1.48913 \pi )$.}
\label{fig:Hist_so31_special}
\end{figure}

Furthermore, it can also be seen in figure~\ref{fig:SO31} that, unlike in the previous Supergravity models, unstable solutions with $\,\Vloc^{(1)} > 0$ exist along the $\,\overline{\textrm{CC'}}\,$ line with $\,\pi < \theta_{\xi} < \frac{3 \pi}{2}$. These solutions require type 1 D6-branes and are compatible with $\,\Vloc^{(2)}<0\,$, so type 2 O6-planes have to be present. The point in the parameter space already studied in section~\ref{sec:Numerical}, in which the two separate moduli solutions of (\ref{2vacua}) coexist, belongs to this set of solution and its sources of potential energy are shown in the left plot of  figure~\ref{fig:Hist_so31_special}. The point in the parameter space having the axion-vanishing moduli VEVs of (\ref{null_axions}), also belong to this class. In this solution,  Re$F_{\Phi}=0\,$ and $\,\,V_{\bar{H}_{3}} = V_{\bar{F}_{4}}=0\,$ together with $\,V_{\bar{F}_{0}} = 0$, as it is displayed in the right plot of figure~\ref{fig:Hist_so31_special}.  

Finally, the flux induced polynomials $\,\cP_{2,3}(\cZ)\,$ in table~\ref{tablecP23} for this Supergravity models, also reduce to those of the \textit{geometric} IIA models in the limit case of $\,\cZ \rightarrow 0$. Therefore, one would expect that, as we approach point B in figure~\ref{fig:SO31}, the profile of the potential energy contributions should match that of figure~\ref{fig:Hist_iso}, again up to a scale factor. Indeed, $\,V_{Q} \rightarrow 0\,$ and $\,V_{R} \rightarrow 0\,$ when we approach this singular point, i.e. $\,|\Phi_{0}| \rightarrow 0\,$, of the moduli VEVs. As for all the previous Supergravity models, dS extrema can again be obtained by continuously deforming the Mkw solutions.

\section{Conclusions}

In this paper we have performed a systematic and complete analysis of the $\mathcal{N}=1$, $D=4$ Supergravity potential induced by generalised fluxes in the context of type II orientifold models that are T-duality invariant and allowed by the symmetries of the  $\,\mathbb{T}^{6}/ (\mathbb{Z}_{2} \times \mathbb{Z}_{2})\,$ isotropic orbifold. The key point throughout this work is the realisation, already presented in refs~\cite{Font:2008vd,deCarlos:2009fq}, that non-geometric $Q$ fluxes, together with NS-NS 3-form $\bar{H}_3$ fluxes, are the structure constants of the Supergravity algebra defined by the isometry and gauge generators that come from the reduction of the metric and the $B$-field. A classification of allowed algebras by the symmetries of the model (including tadpole cancellation conditions arising from the presence of localised sources) was performed in these previous works, and the end result was that there are only five viable Supergravity models. These are parametrised by four real quantities,
($\epsilon_1,\epsilon_2$) that determine the algebra, and ($\xi_3, \xi_7$) that tell us the number and type of localised sources involved.

We have now analysed these five possible models, which result in five different polynomial forms for the superpotential determining the $\mathcal{N}=1$, $D=4$ scalar potential, in order to see whether any of them can contain minima that have all moduli stabilised and Supersymmetry broken. This is a programme that had already begun in ref.~\cite{deCarlos:2009fq}, where we made use of the so-called no-go theorems on the existence of Minkowski/de Sitter vacua already published in the literature. That allowed us to single out the $B$-field reduction based on the semisimple $\,\mathfrak{so(3,1)}\,$ algebra as the promising case that evaded all conditions posed by the no-go theorems.

Technically speaking, the analysis of the extrema of the scalar potential was performed both analytically and numerically. The extremisation with respect to the fields that enter the superpotential linearly (i.e. the dilaton $\cS$ and modulus $\cT$)  is performed analytically, and the resulting solutions, functions of the modulus $\cZ$, are plugged back into $V$. This results in an extremely involved polynomial functions of high powers of $\cZ$, which we solve numerically. The Minkowski condition, $V=0$ is imposed to facilitate the analysis, as well as being of physical interest.

As stated above, only one choice of $B$-field reduction, that based on the $\,\mathfrak{so(3,1)}\,$ algebra, gives rise to minima with all moduli stabilised at a Minkowski vacuum. These solutions can also be deformed continuously to either de Sitter or anti de Sitter by a slight variation of the relevant parameters. Supersymmetry is broken by all moduli, at a scale which is, as expected, large for values of the fluxes of order one. Our systematic search showed that \textit{all} the $B$-field reductions (but the $\mathfrak{nil}$ based one) produce Minkowski extrema with all but one direction stabilised. These {\em tachyonic} solutions show a specific pattern, as they always interpolate between singular points of the parameter space where one or several moduli go to either zero or infinity.
We have also shown the breakdown of the potential energy contributions in the language of type IIA, in order to compare our results to those examples put forward in the context of the no-go theorems. In this way it is obvious that the solutions with stable, Minkowski vacua require non-geometric flux contributions to the scalar potential.

Finally, we would like to make a comment on the applicability of the techniques developed here. The analysis presented can be certainly be performed for a different type of construction. In particular asymmetric orbifolds could be studied, taking as a starting point the results obtained here for symmetric ones.

\vspace*{5mm}
\noindent
{\bf \large Acknowledgments}
\vspace*{3mm}

We are grateful to P.~C\'amara, G.~Dibitetto, A.~Font, D.~Roest, G.~Villadoro, G.~Weatherill and T.~Wrase for useful comments and discussions. A.G. acknowledges the financial support of a FPI (MEC) grant reference BES-2005-8412. This work has been partially supported by CICYT, Spain, under contract FPA 2007-60252, the European Union through the Marie Curie Research Training Network ``UniverseNet'' (MRTN-CT-2006-035863) and the Comunidad de Madrid through Proyecto HEPHACOS S-0505/ESP-0346. The work of BdC is supported by STFC (UK).


\begin{thebibliography}{98}

%
%

\bibitem{deCarlos:2009fq}
  B.~de Carlos, A.~Guarino and J.~M.~Moreno,
  ``Flux moduli stabilisation, Supergravity algebras and no-go theorems,''
  arXiv:0907.5580 [hep-th].


%
%


\bibitem{Hertzberg:2007wc}
  M.~P.~Hertzberg, S.~Kachru, W.~Taylor and M.~Tegmark,
  ``Inflationary Constraints on Type IIA String Theory,''
  JHEP {\bf 0712} (2007) 095
  [arXiv:0711.2512 [hep-th]].

\bibitem{Silverstein:2007ac}
  E.~Silverstein,
  ``Simple de Sitter Solutions,''
  Phys.\ Rev.\  D {\bf 77} (2008) 106006
  [arXiv:0712.1196 [hep-th]].

\bibitem{Haque}
  S.~S.~Haque, G.~Shiu, B.~Underwood and T.~Van Riet,
  ``Minimal simple de Sitter solutions,''
  arXiv:0810.5328 [hep-th] ;
  \\
  U.~H.~Danielsson, S.~S.~Haque, G.~Shiu and T.~Van Riet,
  ``Towards Classical de Sitter Solutions in String Theory,''
  arXiv:0907.2041 [hep-th].

\bibitem{Caviezel:2008tf}
  C.~Caviezel, P.~Koerber, S.~Kors, D.~Lust, T.~Wrase and M.~Zagermann,
  ``On the Cosmology of Type IIA Compactifications on SU(3)-structure Manifolds,''
  arXiv:0812.3551 [hep-th].

\bibitem{Flauger:2008ad}
  R.~Flauger, S.~Paban, D.~Robbins and T.~Wrase,
  ``On Slow-roll Moduli Inflation in Massive IIA Supergravity with Metric Fluxes,''
  arXiv:0812.3886 [hep-th].



%
%


\bibitem{STW}
  J.~Shelton, W.~Taylor and B.~Wecht,
  ``Nongeometric Flux Compactifications,''
  JHEP {\bf 0510} (2005) 085
  [arXiv:hep-th/0508133] ;
  \\
  J.~Shelton, W.~Taylor and B.~Wecht,
  ``Generalized flux vacua,''
  JHEP {\bf 0702}, 095 (2007)
  [arXiv:hep-th/0607015].

\bibitem{Aldazabal:2006up}
  G.~Aldazabal, P.~G.~Camara, A.~Font and L.~E.~Ibanez,
  ``More dual fluxes and moduli fixing,''
  JHEP {\bf 0605} (2006) 070
  [arXiv:hep-th/0602089].

\bibitem{Font:2008vd}
  A.~Font, A.~Guarino and J.~M.~Moreno,
  ``Algebras and non-geometric flux vacua,''
  JHEP {\bf 0812} (2008) 050
  [arXiv:0809.3748 [hep-th]].

\bibitem{Guarino:2008ik}
  A.~Guarino and G.~J.~Weatherill,
  ``Non-geometric flux vacua, S-duality and algebraic geometry,''
  JHEP {\bf 0902} (2009) 042
  [arXiv:0811.2190 [hep-th]].

\bibitem{Aldazabal:2008zza}
  G.~Aldazabal, P.~G.~Camara and J.~A.~Rosabal,
  ``Flux algebra, Bianchi identities and Freed-Witten anomalies in F-theory compactifications,''
  Nucl.\ Phys.\  B {\bf 814} (2009) 21
  [arXiv:0811.2900 [hep-th]].

%
%

\bibitem{GomezReino:2006dk}
   M.~Gomez-Reino and C.~A.~Scrucca,
   ``Locally stable non-supersymmetric Minkowski vacua in supergravity,''
   JHEP {\bf 0605} (2006) 015
   [arXiv:hep-th/0602246].

\bibitem{GomezReino:2006wv}
   M.~Gomez-Reino and C.~A.~Scrucca,
   ``Constraints for the existence of flat and stable non-supersymmetric vacua in supergravity,''
   JHEP {\bf 0609} (2006) 008
   [arXiv:hep-th/0606273].


%
%

\bibitem{Kachru:2003aw}
  S.~Kachru, R.~Kallosh, A.~Linde and S.~P.~Trivedi,
  ``De Sitter vacua in string theory,''
  Phys.\ Rev.\  D {\bf 68} (2003) 046005
  [arXiv:hep-th/0301240].


%
%

\bibitem{Villadoro:2005cu}
  G.~Villadoro and F.~Zwirner,
  ``N = 1 effective potential from dual type-IIA D6/O6 orientifolds with general fluxes,''
  JHEP {\bf 0506} (2005) 047
  [arXiv:hep-th/0503169].

\bibitem{Camara:2005dc}
  P.~G.~Camara, A.~Font and L.~E.~Ibanez,
  ``Fluxes, moduli fixing and MSSM-like vacua in a simple IIA orientifold,''
  JHEP {\bf 0509} (2005) 013
  [arXiv:hep-th/0506066].

\bibitem{Grana:2006kf}
  M.~Grana, R.~Minasian, M.~Petrini and A.~Tomasiello,
  ``A scan for new N=1 vacua on twisted tori,''
  JHEP {\bf 0705} (2007) 031
  [arXiv:hep-th/0609124].

\bibitem{Aldazabal:2007sn}
  G.~Aldazabal and A.~Font,
  ``A second look at N=1 supersymmetric AdS$_4$ vacua of type IIA supergravity``,
  JHEP {\bf 0802} (2008) 086
  [arXiv:0712.1021 [hep-th]].

\bibitem{Caviezel:2008ik}
  C.~Caviezel, P.~Koerber, S.~Kors, D.~Lust, D.~Tsimpis and M.~Zagermann,
  ``The effective theory of type IIA AdS4 compactifications on nilmanifolds and cosets,''
  Class.\ Quant.\ Grav.\  {\bf 26} (2009) 025014
  [arXiv:0806.3458 [hep-th]].

%
%


\bibitem{Derendinger}
  J.~P.~Derendinger, C.~Kounnas, P.~M.~Petropoulos and F.~Zwirner,
  ``Superpotentials in IIA compactifications with general fluxes,''
  Nucl.\ Phys.\  B {\bf 715} (2005) 211
  [arXiv:hep-th/0411276] ;
  \\
  J.~P.~Derendinger, C.~Kounnas, P.~M.~Petropoulos and F.~Zwirner,
  ``Fluxes and gaugings: N = 1 effective superpotentials,''
  Fortsch.\ Phys.\  {\bf 53} (2005) 926
  [arXiv:hep-th/0503229].

\bibitem{Roest:2009dq}
  D.~Roest,
  ``Gaugings at angles from orientifold reductions,''
  Class.\ Quant.\ Grav.\  {\bf 26} (2009) 135009
  [arXiv:0902.0479 [hep-th]].

\bibitem{Dall'Agata:2009gv}
  G.~Dall'Agata, G.~Villadoro and F.~Zwirner,
  ``Type-IIA flux compactifications and N=4 gauged supergravities,''
  arXiv:0906.0370 [hep-th].

\bibitem{Avramis:2009xi}
  S.~D.~Avramis, J.~P.~Derendinger and N.~Prezas,
  ``Conformal chiral boson models on twisted doubled tori and non-geometric string vacua,''
  arXiv:0910.0431 [hep-th].




\end{thebibliography}
\end{document}